\documentclass[lettersize,journal]{IEEEtran}
\usepackage{amsmath,amssymb,amsfonts,amsthm}
\usepackage{algorithmic}
\usepackage{algorithm}
\usepackage{array}
\usepackage{arydshln}
\usepackage[caption=false,font=normalsize,labelfont=sf,textfont=sf]{subfig}
\usepackage{textcomp}
\usepackage{stfloats}
\usepackage{url}
\usepackage{color}
\usepackage{verbatim}
\usepackage{graphicx}
\usepackage{cite}

\usepackage{hyperref}
\usepackage[normalem]{ulem}
\usepackage{multirow}
\usepackage{enumitem}
\usepackage{float}
\usepackage{caption}
\usepackage{subcaption}
\usepackage{makecell}
\usepackage{enumerate}

\usepackage{xspace}
\newcommand{\etal}{et al.\xspace}
\newcommand{\eg}{e.g.,\xspace}
\newcommand{\ie}{i.e.,\xspace}

\newcommand{\model}{\emph{HLoG}\xspace}

\begin{document}

%
\title{Hierarchical Local-Global Feature Learning for Few-shot Malicious Traffic Detection}


        
        

\author{
Songtao Peng, 
Lei Wang,
Wu Shuai,
Hao Song,
Jiajun Zhou,
Shanqing Yu,
Qi Xuan,~\IEEEmembership{Senior Member,~IEEE}

\IEEEcompsocitemizethanks{
\IEEEcompsocthanksitem 
This work was supported in part by China Post-Doctoral Science Foundation under Grant 2024M762912, in part by the Post-Doctoral Science Preferential Funding of Zhejiang Province of China under Grant ZJ2024060, in part by the Key Research and Development Program of Zhejiang under Grant 2022C01018, in part by the National Natural Science Foundation of China under Grant U21B2001, and in part by the Baima Lake Laboratory Joint Fund of Zhejiang Provincial Natural Science Foundation of China under Grant LBMHZ25F020002.
\emph{(Corresponding author: Jiajun Zhou.)}
\IEEEcompsocthanksitem S. Peng, L. Wang, W. Shuai, H. Song, S. Yu, Q. Xuan are with the Institute of Cyberspace Security, Zhejiang University of Technology, Hangzhou 310023, China, with the Binjiang Institute of Artificial Intelligence, ZJUT, Hangzhou 310056, China. E-mail: pengst@zjut.edu.cn.
\IEEEcompsocthanksitem J. Zhou are with the Institute of Cyberspace Security, College of Computer Science and Technology, Zhejiang University of Technology, Hangzhou 310023, China, with the Binjiang Institute of Artificial Intelligence, ZJUT, Hangzhou 310056, China. E-mail: jjzhou@zjut.edu.cn.
}}



\maketitle

\begin{abstract}
  With the rapid growth of internet traffic, malicious network attacks have become increasingly frequent and sophisticated, posing significant threats to global cybersecurity. Traditional detection methods, including rule-based and machine learning-based approaches, struggle to accurately identify emerging threats, particularly in scenarios with limited samples. While recent advances in few-shot learning have partially addressed the data scarcity issue, existing methods still exhibit high false positive rates and lack the capability to effectively capture crucial local traffic patterns. In this paper, we propose \model, a novel hierarchical few-shot malicious traffic detection framework that leverages both local and global features extracted from network sessions. \model employs a sliding-window approach to segment sessions into phases, capturing fine-grained local interaction patterns through hierarchical bidirectional GRU encoding, while simultaneously modeling global contextual dependencies. We further design a session similarity assessment module that integrates local similarity with global self-attention-enhanced representations, achieving accurate and robust few-shot traffic classification. Comprehensive experiments on three meticulously reconstructed datasets demonstrate that \model significantly outperforms existing state-of-the-art methods. Particularly, \model achieves superior recall rates while substantially reducing false positives, highlighting its effectiveness and practical value in real-world cybersecurity applications. 
\end{abstract}

\begin{IEEEkeywords}
Malicious Traffic Detection, Few-shot Learning, Session Classification, Cybersecurity
\end{IEEEkeywords}

\section{Introduction}
With the rapid advancement of internet technologies, networks have penetrated every aspect of societal life, evolving into a critical global infrastructure. As of December 2022, the global internet user base reached approximately 5.54 billion, with a penetration rate of 69\%~\cite{2}. This unprecedented growth has driven the diversification and complexity of network traffic, propelling global digital transformation and societal efficiency while simultaneously exposing significant vulnerabilities in cyberspace. Currently, internet ecosystems face increasingly frequent malicious activities, with cyberattacks growing in both frequency and severity. These sophisticated threats, often highly covert and resistant to rapid detection, have the potential to inflict substantial economic losses and societal disruptions once deployed. Consequently, the efficient and accurate detection and defense against malicious traffic have emerged as critical research challenges in cybersecurity.

Existing approaches for malicious traffic detection primarily fall into three categories: rule-based, machine learning-based, and deep learning-based methods. Rule-based methods~\cite{3,4,5,6}, such as Snort and Zeek, generally match predefined rules against specific traffic fields to identify malicious activities. Despite their efficiency in identifying known attack patterns, these methods exhibit limited adaptability to novel and encrypted traffic, leading to high false negatives. Machine learning-based methods~\cite{7,8,9,10,11,12} employ manually engineered features combined with classifiers to enhance detection accuracy. However, their effectiveness is heavily dependent on feature quality and can degrade rapidly with evolving attack patterns. Deep learning-based methods~\cite{13,14,15,16,17,18,19,20,21,22}, which automatically learn representations directly from raw data, provide superior adaptability and accuracy. Nevertheless, deep learning models often require large-scale labeled datasets and struggle in scenarios with limited samples, such as zero-day attacks.
Moreover, the emergence of new types of malicious traffic necessitates retraining of deep learning models, thereby constraining their ability to promptly adapt to emerging threats.

To address the aforementioned shortcomings, few-shot learning techniques have been gradually introduced into the cybersecurity domain, aiming to overcome the poor generalization performance of existing methods when limited samples are available. For instance, Xu \etal\cite{27} proposed a meta-learning framework that constructs both a feature extraction network and a sample comparison network to identify unknown traffic from only a few attack samples. Similarly, Feng \etal\cite{28} combined statistical features with packet dependency features and employed a meta-learning model to train a few-shot malicious traffic detection task. Despite these advancements, current few-shot malicious traffic detection methods still face two critical challenges: 1) the detection models exhibit high false positive rates, often misclassifying a substantial amount of benign traffic as malicious, thereby undermining their practical applicability; 2) existing approaches lack effective mechanisms to capture the local details of network traffic, overlooking subtle yet crucial attack features that are indicative of malicious behavior, which hampers the accurate identification of novel attack traffic.

\begin{figure}[!htb]
  \centering
  \includegraphics[width=\linewidth]{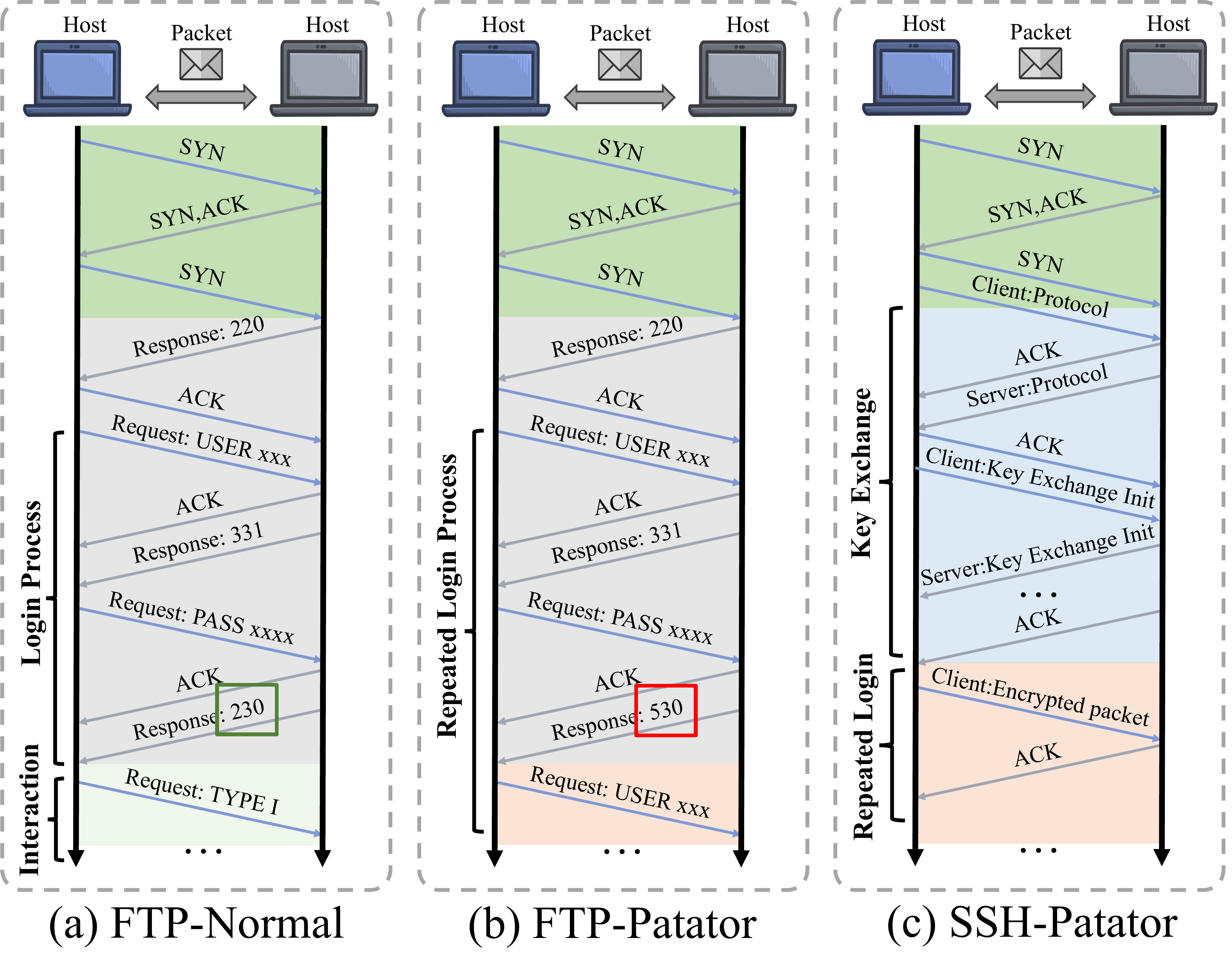}
  \caption{Example of different network traffic interaction scenarios.}
  \label{fig: example}
\end{figure}

In fact, network traffic contains a wealth of local details that reflect attack characteristics, and these details often play a critical role in detection. As illustrated in Fig.~\ref{fig: example}, we present three distinct traffic interaction scenarios: 
(a) normal FTP traffic, where a TCP connection is established followed by a standard username and password verification before proceeding with regular data exchange; 
(b) FTP-Patator attacks, characterized by repeated username and password verification attempts aimed at cracking login credentials; 
and (c) SSH-Patator attacks, while also establishing a TCP connection, involve a key exchange due to the encrypted nature of the protocol, followed by multiple repeated attempts at username and password verification. 
Analyzing these examples reveals that the initial phase (i.e., TCP connection establishment) is nearly identical across scenarios, whereas the subsequent interaction patterns (such as repeated authentication attempts) exhibit significant local differences. This indicates that, in malicious traffic detection, if a model can effectively segment network traffic into its distinct interaction phases and accurately capture the local details, it will be more capable of enhancing detection accuracy and reducing false positive rates. Nonetheless, several challenges remain, including how to appropriately partition the local phases of network traffic, how to precisely compute and integrate local and global features, and how to evaluate the similarity between different sessions under few-shot conditions.

To address these challenges, we propose \model , a few-shot malicious traffic detection method based on \textbf{H}ierarchical \textbf{Lo}cal-\textbf{G}lobal feature learning. The approach first segments traffic sessions using a sliding window mechanism and employs a hierarchical feature encoding strategy to capture localized and global traffic characteristics. Subsequently, cross-phase local similarity computation and global self-attention enhancement are leveraged to assess traffic similarities, enabling precise characterization of subtle inter-traffic differences. This method achieves significant performance improvements in few-shot attack detection, particularly excelling in reducing false positives and enhancing recall.
The main contributions of this paper are summarized as follows:
\begin{itemize}[leftmargin=10pt]
  \item We systematically analyze and reorganize multiple public malicious traffic datasets to reconstruct three category-rich datasets tailored for few-shot scenarios. These new datasets address critical limitations of existing data (\eg limited categories, class imbalance) and provide standardized foundations for few-shot malicious traffic detection research.
  \item We propose \model, a novel hierarchical few-shot malicious traffic detection framework that simultaneously captures and fuses localized fine-grained features and global contextual features. This dual focus enables precise similarity measurement between traffic samples, significantly enhancing generalization to novel attack types.
  \item Comprehensive evaluations across multiple datasets demonstrate that \model achieves state-of-the-art performance, particularly excelling in reducing false positive rates and improving recall. These results supported by ablation studies highlight the model's effectiveness and practical applicability in real-world cybersecurity scenarios.
\end{itemize}





\section{Related Work} \label{sec:related work}
\subsection{Traditional Malicious Traffic Detection}
\subsubsection{Rule-based Detection Methods} 
Rule-based detection methods are among the simplest and most efficient techniques for malicious traffic detection, known for their high accuracy and real-time capabilities, making them well-suited for deployment in high-bandwidth network environments~\cite{29}. These methods identify malicious activity by matching observed network traffic against a predefined set of rules. Prominent examples include Snort~\cite{3}, which classifies packets based on rules applied to packet headers and payloads, and Zeek (formerly Bro)~\cite{5}, which converts network traffic into high-level events for subsequent analysis via script-based policy interpreters. While such methods typically exhibit low false positive rates when correctly configured, they heavily rely on predefined rules formulated from known attack signatures. Consequently, they struggle to detect novel, zero-day attacks~\cite{30}. Furthermore, with the increasing prevalence of encrypted traffic and the evolving complexity of network environments, rule-based systems face significant limitations, resulting in increased false negative rates.

\subsubsection{Machine Learning-based Detection Methods} 
Advances in machine learning have led to widespread adoption of these techniques for malicious traffic detection tasks~\cite{31,32,33}. Early efforts primarily involved supervised machine learning algorithms applied to manually crafted features~\cite{7,8,9}. Recently, unsupervised learning has garnered increased attention due to its ability to detect unknown attacks without labeled training data. For example, Mirsky \etal~\cite{12} proposed Kitsune, a method that uses a stacked autoencoder architecture for unsupervised extraction and classification of network traffic features. Fu \etal~\cite{29} developed Whisper, which extracts frequency-domain features from packet sequences for clustering-based malicious traffic detection. Although machine learning-based methods substantially reduce false negative rates compared to rule-based approaches, their performance strongly depends on the quality of manually engineered features. This dependence makes them less adaptable to new or rapidly evolving traffic patterns and often necessitates considerable feature engineering efforts to maintain detection accuracy in diverse network contexts.

\subsubsection{Deep Learning-based Detection Methods} 
Deep learning approaches have significantly advanced malicious traffic detection capabilities by automating feature extraction from raw network data, thus overcoming the limitations of manually designed features inherent to traditional machine learning methods. Several works have utilized raw packet data directly as input to deep neural networks to learn discriminative traffic representations. For instance, Holland \etal~\cite{16} proposed encoding packets using normalized binary representations, preserving semantic structures, and leveraging AutoML to construct packet-level classification models. Xiao \etal~\cite{17} introduced the Extended Byte Segment Neural Network (EBSNN), designed explicitly for network traffic classification tasks. Furthermore, deep learning has also been successfully employed to extract temporal features from sequential network data. Cui \etal~\cite{18} designed a multi-sequence fusion classifier based on Transformer architectures to identify malware-related traffic behaviors. Additionally, recent developments in graph neural networks (GNNs)~\cite{kipf2016semi,zhou2025clarify} have further enhanced the capacity to model complex network traffic interactions. Shen \etal~\cite{20} and Hu \etal~\cite{21} proposed modeling session packet sequences as Traffic Interaction Graphs (TIG), subsequently leveraging graph-based representation learning and classification techniques for improved detection performance.
Song \etal~\cite{hao2024network} proposed MuFF, a multi-view feature fusion method that captures both temporal and interactive packet-level relationships to address limitations of single-view detection models, demonstrating superior performance on multiple real-world datasets.
Zhou \etal~\cite{zhou2025multi} introduced FlowID, a multi-view correlation-aware framework that combines temporal and interaction-aware traffic modeling with hypergraph representation learning and dual-contrastive proxy tasks, significantly improving detection accuracy and robustness against data imbalance and label scarcity in diverse network scenarios.
Although deep learning-based approaches achieve superior accuracy and generalization through deeper architectures and automated feature extraction, their effectiveness strongly depends on large-scale labeled datasets. Consequently, in scenarios characterized by limited labeled data, these methods are prone to overfitting, thus necessitating specialized strategies, such as data augmentation~\cite{34,wu2024eg,zhou2022behavior}, transfer learning, or few-shot learning, to achieve robust detection performance.

\subsection{Few-shot Malicious Traffic Detection}
To overcome the inherent limitation of deep learning methods requiring extensive labeled samples, researchers have recently explored few-shot learning techniques to improve malicious traffic detection performance under limited-data conditions. Current research on few-shot malicious traffic detection can generally be categorized into two strategies: data-based methods and model-based methods.

\subsubsection{Data-based Methods}
Data-based methods aim to alleviating data imbalance data scarcity issues via undersampling, oversampling, and data augmentation techniques~\cite{34}. For instance, Al and Dener~\cite{35} proposed a hybrid sampling strategy combining SMOTE oversampling with Tomek-Links undersampling, training convolutional neural network (CNN) and long short-term memory (LSTM) classifiers to enhance malicious traffic detection. Additionally, generative adversarial networks (GANs), well-known for their data augmentation capabilities due to their strong ability to model complex data distributions~\cite{39}, have been extensively applied to malicious traffic detection for generating synthetic minority-class samples~\cite{37,38,39,40}. However, these methods often face challenges related to the limited number of available minority-class samples, making it difficult to generate realistic, high-quality synthetic data. Furthermore, the interpretability and trustworthiness of GAN-generated samples are difficult to validate, posing significant barriers for practical deployment~\cite{26}.

\subsubsection{Model-based Approaches}
Model-based methods, on the other hand, employ few-shot learning algorithms directly within the model architectures, typically encompassing two main strategies: model fine-tuning and transfer learning. Model fine-tuning approaches usually begin with a large-scale pretraining phase, subsequently fine-tuning on limited labeled samples to achieve accurate detection. For example, Lin \etal~\cite{41} pretrained a traffic representation model using large unlabeled datasets and fine-tuned it on limited labeled data to classify encrypted traffic effectively. Transfer learning methods leverage task sampling strategies to simulate novel scenarios during training, facilitating rapid model adaptation to new types of malicious traffic. Xu \etal~\cite{27} proposed a meta-learning framework capable of distinguishing between normal and malicious traffic using a few-shot learning paradigm. Similarly, Feng \etal~\cite{28} aggregated statistical and packet sequence features to form representation vectors, subsequently employing model-agnostic meta learning (MAML) to achieve efficient few-shot malicious traffic detection.

Although existing model-based methods have demonstrated promising results, significant challenges remain. For example, fine-tuning approaches tend to overfit severely when the target scenarios differ significantly from training scenarios. Additionally, current transfer learning-based approaches often focus on binary intrusion detection tasks (benign vs. malicious) without exploring more detailed classification of diverse malicious traffic subcategories. Moreover, prior works have not thoroughly investigated suitable feature representation methods specifically tailored for few-shot learning scenarios, underscoring the need for further advancements in model architectures and feature extraction strategies.

\begin{table}[!htb]
  \renewcommand\arraystretch{1.2}
  \caption{Main notations used in this paper.}
  \centering
  \resizebox{\linewidth}{!}{
  \begin{tabular}{lr}
  \hline\hline
  Notation                                      & Definition          \\ \hline
  $\mathcal{D}=\{(x_i,y_i),\cdots\}$           & $\text{Dataset}=\{(\text{sample}, \text{label}),\cdots\}$             \\
  $\mathcal{C},\mathcal{C}^\text{mal}$   & label set, label set of malicious classes            \\
  $M$     & number of malicious calsses\\
  $T,\mathcal{T},\mathcal{S},\mathcal{Q}$       & episode, episode set, support set, query set             \\
  $J,K$                                     & the number of classes and samples per class in support set.              \\
  $\boldsymbol{s},\boldsymbol{S},\boldsymbol{P}$        & packet sequence, session representation, session phase            \\
  $N,Q$                                       & number of packets in a session, window length \\
  $\boldsymbol{h},\boldsymbol{z}$ & hidden state, feature representation of session phase              \\
  $\boldsymbol{Z}_\text{local}, \boldsymbol{Z}_\text{global}$    & local feature and global feature of a session   \\
  $d,l$                         & number of hidden units in GRU, number of layers in Bi-GRU               \\
  $\boldsymbol{L}$        & local similarity matrix                \\ 
  $\boldsymbol{y}$     & one-hot encoded label    \\
  \hline\hline
  \end{tabular}}
  \label{tab: notation}
\end{table}

\subsection{Summary and Motivation of Our Work}
In summary, while both traditional and recent few-shot malicious traffic detection methods have made significant progress, several critical challenges remain unsolved. Existing approaches frequently suffer from high false positive rates due to limited generalization and insufficient attention to capturing detailed, local traffic features, resulting in reduced accuracy in detecting novel or subtle attacks. Addressing these gaps, this work aims to develop a hierarchical few-shot malicious traffic detection method capable of effectively capturing both local and global features, providing precise similarity assessments between traffic sessions, and significantly improving detection accuracy and robustness under few-shot conditions.

\section{Preliminaries}
\subsection{Network Traffic}
\begin{figure}[!htb]
  \centering
  \includegraphics[width=\linewidth]{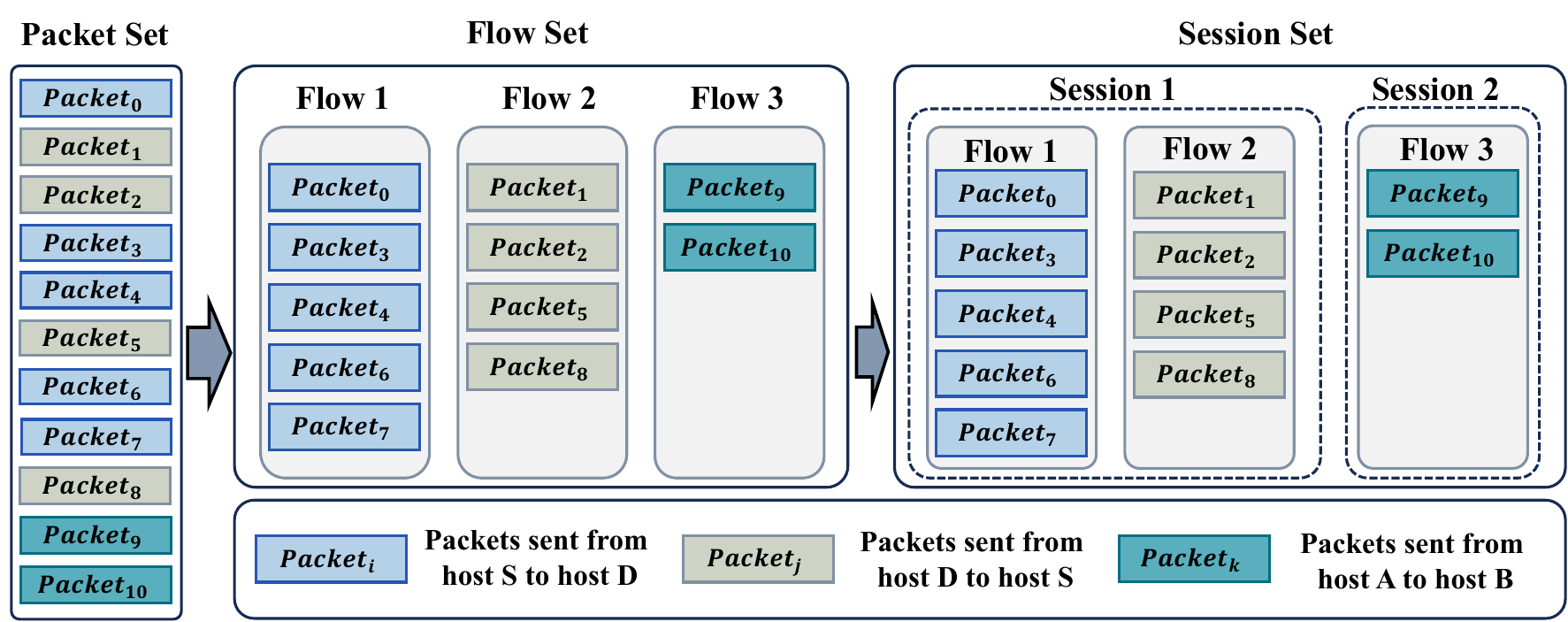}
  \caption{An example illustrating the relationship among packets, flows, and sessions in network traffic.}
  \label{fig: session}
\end{figure}

In real-world network environments, the minimal transmission unit is a data packet composed of a header and a payload. The header contains critical information such as source IP address, destination IP address, and port numbers. Packets sharing the same five-tuple (source IP, destination IP, source port, destination port, and transport layer protocol) collectively constitute a flow. When bidirectional data transmission occurs between two endpoints, each communicating party generates an independent flow. \textit{The combination of these two corresponding unidirectional flows constitutes a \textbf{session}}. 
Within this hierarchical structure, a session not only preserves the complete interactive process between communication endpoints but also effectively characterizes temporal features and protocol interaction patterns of network behaviors. Consequently, \textbf{this work establishes the session as the minimal analytical unit for malicious traffic detection, \ie detecting malicious traffic through session-level classification}. As illustrated in Fig.~\ref{fig: session}, the original captured data comprising 11 packets is clustered based on the five-tuple into three distinct flows, and Session 1, formed by combining bidirectional flows, represents a typical detection sample instance.

\subsection{Different Paradigms for Malicious Traffic Detection}\label{sec:problem}
In this subsection, we provide a detailed introduction to the methodologies and dataset construction strategies for traditional malicious traffic detection and few-shot malicious traffic detection.
Consider a dataset containing $n$ network traffic samples defined as $\mathcal{D}=\{(x_1, y_1), (x_2, y_2), \dots, (x_n, y_n)\}$, where each sample $x_i$ represents a network traffic session. The corresponding labels $y_i \in \mathcal{C}=\{0, 1, 2, \dots, M\}$ indicate the traffic type, where $y_i = 0$ denotes benign (normal) traffic and $y_i \in \mathcal{C}^\text{mal}=\{1, 2, \dots, M\}$ denotes different malicious traffic types, with $M$ being the total number of malicious traffic categories in the dataset.
\begin{figure}[!htb]
  \centering
  \includegraphics[width=\linewidth]{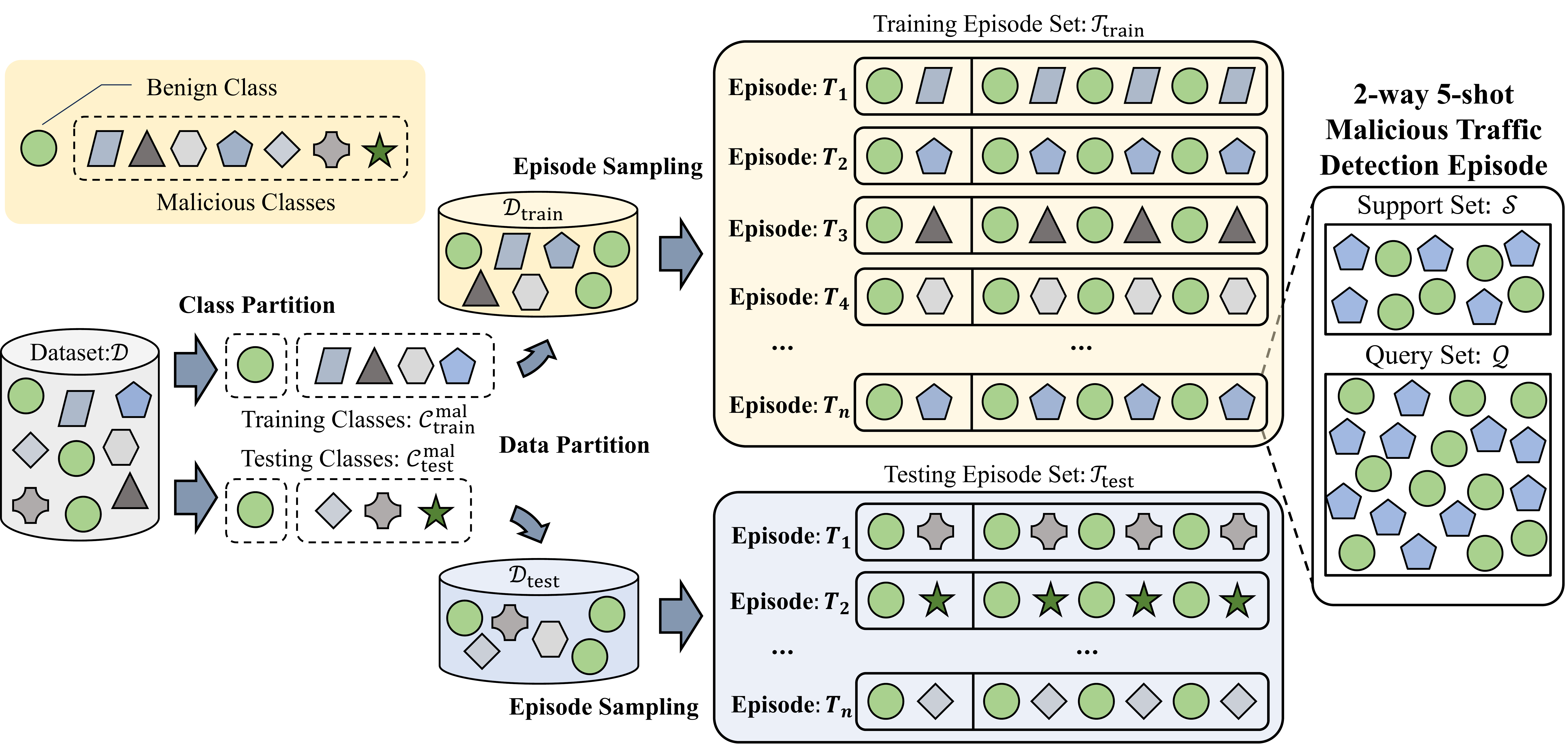}
  \caption{Illustration of data partitioning and episode construction in few-shot learning settings.}
  \label{fig: few-shot}
\end{figure}
\subsubsection{Traditional Malicious Traffic Detection}
In traditional scenarios, individual samples serve as the basic units for model training and evaluation. Typically, the dataset $\mathcal{D}$ is partitioned into non-overlapping training set $\mathcal{D}_\text{train}$ and testing set $\mathcal{D}_\text{test}$ based on a predefined ratio. The detection model $f(x_i)\mapsto y_i$ is trained using $\mathcal{D}_\text{train}$, enabling it to predict the corresponding labels from input samples. This paradigm relies on sufficient training data to accurately capture traffic features.

\subsubsection{Few-shot Malicious Traffic Detection}
Unlike traditional paradigm, few-shot malicious traffic detection emphasizes improving model generalization under limited-sample conditions. In this setting, the basic analytical unit is no longer individual samples, but rather \textbf{episode}. Each episode $T$ comprises a support set $\mathcal{S}$, providing class information, and a query set $\mathcal{Q}$, which evaluates the model's classification capabilities. An episode is defined as an ``$J$-way, $K$-shot'' detection problem if the support set consists of $J$ classes with only $K$ samples per class. The detailed procedure is shown in Fig.~\ref{fig: few-shot}.
\begin{itemize}[leftmargin=10pt]
  \item \textbf{Class Partitioning:} First, the malicious traffic classes in the dataset $\mathcal{D}$ are split into two non-overlapping subsets: training classes $\mathcal{C}^\text{mal}_\text{train}$ and testing classes $\mathcal{C}^\text{mal}_\text{test}$.
  \item \textbf{Sub-dataset Construction:} Malicious samples corresponding to $\mathcal{C}^\text{mal}_\text{train}$ and $\mathcal{C}^\text{mal}_\text{test}$ are extracted from $\mathcal{D}$ to form malicious training and testing subsets $\mathcal{D}^\text{mal}_\text{train}$ and $\mathcal{D}^\text{mal}_\text{test}$, respectively. Similarly, benign traffic (label 0) is also partitioned accordingly to form corresponding training and testing subsets: $\mathcal{D}^\text{ben}_\text{train}$ and $\mathcal{D}^\text{ben}_\text{test}$.
  \item \textbf{Training and Testing Set Formation}: The final training set $\mathcal{D}_\text{train}$ is formed by combining $\mathcal{D}^\text{mal}_\text{train}$ and $\mathcal{D}^\text{ben}_\text{train}$, while the testing set $\mathcal{D}_\text{test}$ is formed by combining $\mathcal{D}^\text{mal}_\text{test}$ and $\mathcal{D}^\text{ben}_\text{test}$, \ie $\mathcal{D}_\text{train}=\mathcal{D}^\text{mal}_\text{train}\cup\mathcal{D}^\text{ben}_\text{train}$ and $\mathcal{D}_\text{test}=\mathcal{D}^\text{mal}_\text{test}\cup\mathcal{D}^\text{ben}_\text{test}$.
  \item \textbf{Episode Sampling}: From $\mathcal{D}_\text{train}$, multiple training episodes $\mathcal{T}_\text{train} = \{T_1, T_2, \dots\}$ are randomly sampled, each comprising a support set and a query set. Similarly, testing episodes $\mathcal{T}_\text{test}$ are sampled from $\mathcal{D}_\text{test}$.
\end{itemize}
Through episodic training, the model $f$ acquires generalized meta-knowledge about malicious traffic, enabling accurate and efficient classification of query samples given limited support data, and achieving strong performance on test tasks.

\begin{figure*}[!htb]
  \centering
  \includegraphics[width=\textwidth]{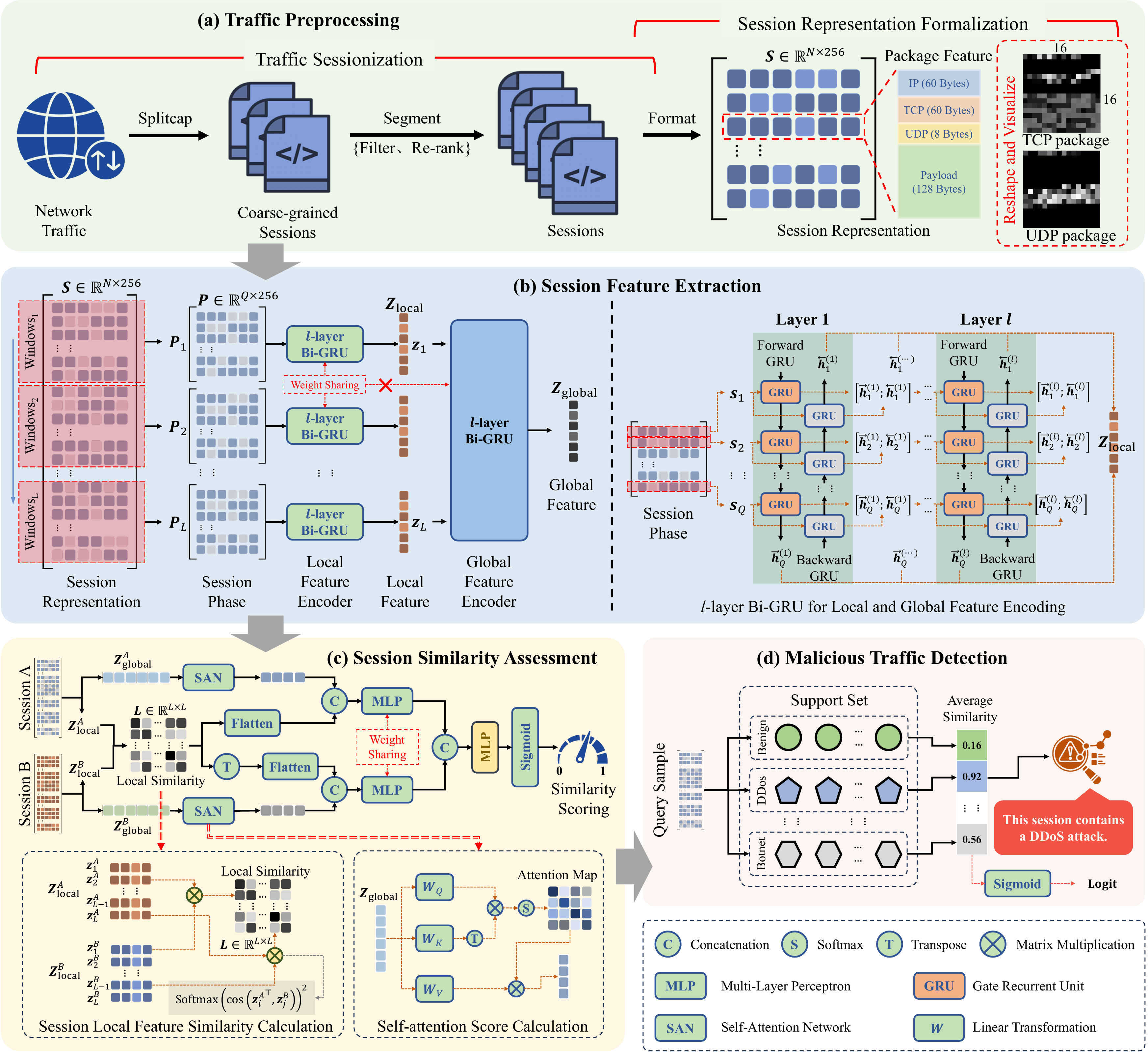}
  \caption{Illustration of the \model framework. The complete workflow proceeds as follows: 
  (1) converting raw traffic into standardized session representations; 
  (2) extracting hierarchical session features via local interaction modeling and global dependency encoding; 
  (3) computing cross-session similarities by fusing local similarity and self-attention enhanced global features; 
  (4) determining malicious traffic through similarity-based few-shot classification.}
  \label{fig: framework}
\end{figure*}



\section{Methodology}\label{sec: method}
In this section, we detail the proposed framework \model for few-shot malicious traffic detection, as schematically illustrated in Fig.~\ref{fig: framework}. \model takes raw network traffic as input and outputs malicious traffic classification results through four core components: 
(1) a \textbf{traffic preprocessing module} that converts raw packets into standardized session representations with temporal integrity and anonymized features;
(2) a \textbf{hierarchical session feature extraction module} that employs bidirectional GRU networks to learn both local interaction patterns and global session semantics through hierarchical encoding;
(3) a \textbf{session similarity assessment module} that fuses local-phase correlations and enhanced global features through attention mechanisms for robust similarity learning;
(4) a \textbf{similarity-based classification mechanism} that determines malicious categories by comparing query samples with few-shot support sets.
The framework is optimized end-to-end through a mean squared error objective that enforces discriminative similarity measurements between different traffic categories. We now elaborate on each core component.

\subsection{Traffic Preprocessing}
The traffic preprocessing pipeline aims to convert raw network traffic into standardized session representations with unified structures, thereby providing a reliable data foundation for subsequent malicious traffic detection. Fig.~\ref{fig: framework}(a) illustrates the complete preprocessing workflow.

\subsubsection{Traffic Sessionization}
In network traffic collection, tools such as Wireshark and tcpdump are typically employed to capture packets at network interfaces, storing them as raw traffic files in PCAP format. These raw files are often large in size and contain mixed data from multiple sessions. Since the detection target of this work is individual sessions, preprocessing is required to extract standardized session data suitable for model input. Specifically, we first utilize Splitcap to segment raw traffic into sessions based on the five-tuple criterion, generating independent PCAP files. During this process, irrelevant packets—such as ICMP/ARP packets lacking transport-layer information and retransmitted packets caused by transmission errors—are filtered out to eliminate noise from non-target protocols and network anomalies.

However, session segmentation solely based on five-tuples may result in a single session file containing multiple interaction processes due to port reuse. Consequently, further refinement is required to extract fine-grained session files corresponding to individual interaction processes. Moreover, to address packet sequencing anomalies such as fragmentation and retransmission-induced disorder during network transmission, preprocessing includes reordering out-of-sequence packets and filtering retransmissions. Specifically, differentiated processing is applied based on transport-layer protocols. For TCP sessions, independent and complete interaction cycles are identified by inspecting flag bits in the three-way handshake (SYN, SYN-ACK, ACK) and four-way termination (FIN-ACK), while out-of-order packets are reordered via sequence numbers to ensure temporal integrity. For UDP sessions, which lack connection-oriented mechanisms, interaction cycles are segmented based on time interval between packets: splits occur if the interval between adjacent packets exceeds 60 seconds or if the interval from the first packet in the current interaction exceeds 120 seconds. Additionally, since UDP lacks sequence control, packets cannot be reliably reordered or retransmissions identified; thus, no filtering or reordering is performed. Finally, sessions containing only 1-2 packets are discarded, as these typically include trivial data such as TCP handshake packets or network time protocols, which provide insufficient information and potentially introduce noise.

\subsubsection{Session Representation Formalization}
After sessionization, a standardized dataset comprising multiple complete session samples is obtained. However, these sessions are not immediately suitable as model inputs due to varying packet counts and lengths. Thus, we further format packets and sessions into unified representations, detailed as follows:
\begin{itemize}[leftmargin=10pt]
  \item \textbf{Session Length Alignment:} For each session, the first $N$ packets are retained as feature carriers. In cases where fewer than $N$ packets exist, zero-padded dummy packets are appended to ensure all sessions are standardized to $N$-packet sequences.
  \item \textbf{Packet Anonymization:} To prevent model dependency on specific IP addresses and improve generalization, we anonymize packets by removing MAC addresses from data-link layer. Additionally, the source and destination IP addresses in the first packet of each session are replaced with placeholder addresses 0.0.0.0 and 255.255.255.255, respectively. This approach preserves transmission directionality while obscuring real addresses, aiding detection model in capturing internal session structures.
  \item \textbf{Packet Feature Alignment:} Key fields of each packet are aligned into a unified vector format. Specifically, IP headers are aligned to a maximum length of 60 bytes, with insufficient lengths filled with zeros to form the IP-layer features. For the transport layer, protocol-specific padding is performed. For TCP (protocol number 6), the TCP header is padded to 60 bytes, followed by an 8-byte zero-filled UDP placeholder, totaling 68 bytes. For UDP (protocol number 17), a 60-byte zero-filled TCP placeholder is added before the 8-byte UDP header, also forming a 68-byte transport-layer feature vector. For application-layer payloads, the first 128 bytes are extracted as application-layer features. Consequently, each packet is represented as a 256-byte vector composed of IP-layer (60 bytes), transport-layer (68 bytes), and application-layer (128 bytes) features, as depicted in Fig.~\ref{fig: framework}(a).
  \item \textbf{Representation Normalization:} Each byte value is normalized from the range $[0, 255]$ to $[0, 1]$, enhancing numerical stability and gradient propagation efficiency during model training. The normalized packet sequences $\{\boldsymbol{s}_1, \boldsymbol{s}_2, \cdots, \boldsymbol{s}_N\in\mathbb{R}^{256}\}$ constitute the finalized session representation $\boldsymbol{S}\in\mathbb{R}^{N\times 256}$.
\end{itemize}
Fig.~\ref{fig: framework}(a) visualizes preprocessed packets from TCP and UDP sessions, reshaped into $16\times 16$ grayscale images. These examples highlight distinct feature distribution patterns between protocol-specific packet structures.

\subsection{Session Feature Extraction}
In malicious traffic detection, directly modeling complete sessions poses dual challenges. First, network attacks often manifest as short-term, high-frequency local patterns, which may be diluted by global modeling. Second, long sessions may lead to gradient vanishing issues during modeling, weakening the model's ability to capture long-range dependencies. To enable effective representation learning for network sessions, we propose a hierarchical sequence encoding method. This method first extracts local interaction features from different session phases, then derives global session representations from temporal dependencies across phases, enhancing the characterization of complex traffic behavior patterns. The process is illustrated in Fig.~\ref{fig: framework}(b).

\subsubsection{Local Feature Encoding}
To effectively capture local information in long sessions, we apply a sliding window of length $Q$ to segment each session into multiple phases. Each phase corresponds to a submatrix $\boldsymbol{P}_i \in \mathbb{R}^{Q \times 256}$, allowing the entire session to be represented as a collection of segments: $\boldsymbol{S}=[\boldsymbol{P}_1, \boldsymbol{P}_2, \dots, \boldsymbol{P}_L]$, where $L=\lceil N/Q \rceil$ denotes the number of phases. This segmentation strategy preserves local behavioral integrity while reducing the complexity of long-sequence modeling. Additionally, it allows parallelized feature extraction across phases, significantly improving computational efficiency to meet real-time deployment requirements.
Furthermore, bidirectional semantic dependencies between request and response traffic in network sessions necessitate joint analysis to identify attack intent. To address this, we employ a multi-layer Bidirectional Gated Recurrent Units (Bi-GRU) for local feature extraction. Specifically, the Bi-GRU processes packet sequences in a session phase using forward and backward networks, generating corresponding hidden states:
\begin{equation}
  \begin{aligned}
  &\text{Layer 1}\quad\text{Forward:}\quad\overrightarrow{\boldsymbol{h}}_t^{(1)}=\overrightarrow{\textsf{GRU}}\left(\boldsymbol{s}_t, \overrightarrow{\boldsymbol{h}}_{t-1}^{(1)}\right) \\
  &\text{Layer 1}\quad\text{Backward:}\quad\overleftarrow{\boldsymbol{h}}_t^{(1)}=\overleftarrow{\textsf{GRU}}\left(\boldsymbol{s}_t, \overleftarrow{\boldsymbol{h}}_{t+1}^{(1)}\right) \\
  &\text{Output in step}\ t\text{:}\quad\boldsymbol{h}_t^{(1)}=\left[\overrightarrow{\boldsymbol{h}}_t^{(1)};\overleftarrow{\boldsymbol{h}}_t^{(1)} \right]\in\mathbb{R}^{2d}
  \end{aligned}
\end{equation}
where $\overrightarrow{\boldsymbol{h}}_t^{(1)}$, $\overleftarrow{\boldsymbol{h}}_t^{(1)}\in\mathbb{R}^d$ are the forward and backward hidden states at the $t$-th step of the first layer respectively, $d$ denotes the number of hidden units in the unidirectional GRU. Furthermore, we stack $l$ layers of Bi-GRU to extract high-order features, where lower layers capture local temporal patterns (e.g., single-packet payload anomalies), while higher layers combine low-order features through cross-layer parameter sharing to represent complex behaviors (e.g., multi-packet covert channels):
\begin{equation}
  \begin{aligned}
  &\text{Layer $l$}\quad\text{Forward:}\quad\overrightarrow{\boldsymbol{h}}_t^{(l)}=\overrightarrow{\textsf{GRU}}\left(\boldsymbol{h}_t^{(l-1)}, \overrightarrow{\boldsymbol{h}}_{t-1}^{(l)}\right) \\
  &\text{Layer $l$}\quad\text{Backward:}\quad\overleftarrow{\boldsymbol{h}}_t^{(l)}=\overleftarrow{\textsf{GRU}}\left(\boldsymbol{h}_t^{(l-1)}, \overleftarrow{\boldsymbol{h}}_{t+1}^{(l)}\right) \\
  &\text{Output in timestep}\ t\text{:}\quad\boldsymbol{h}_t^{(l)}=\left[\overrightarrow{\boldsymbol{h}}_t^{(l)};\overleftarrow{\boldsymbol{h}}_t^{(l)} \right]
  \end{aligned}
\end{equation}
Each session phase $\boldsymbol{P}=\left[\boldsymbol{s}_1, \boldsymbol{s}_2, \cdots, \boldsymbol{s}_Q\right]$ is independently input into the local feature encoder, yielding a representation vector $\boldsymbol{z}=\textsf{LF-Bi-GRU}(\boldsymbol{P}_i)$ as follows:
\begin{equation}
  \boldsymbol{z}=\left[\overrightarrow{\boldsymbol{h}}_Q^{(1)},\overleftarrow{\boldsymbol{h}}_1^{(1)},\cdots,\overrightarrow{\boldsymbol{h}}_Q^{(l)},\overleftarrow{\boldsymbol{h}}_1^{(l)}\right]\in \mathbb{R}^{2\cdot l \cdot d}
\end{equation}
The representations of all session phases are then concatenated to form the session's local feature sequence:
\begin{equation}
  \boldsymbol{Z}_\text{local}=\left[\boldsymbol{z}_1, \boldsymbol{z}_2,\cdots,\boldsymbol{z}_L\right]
\end{equation}
\subsubsection{Global Feature Encoding}
After obtaining local session features, we further capture high-order semantic information of the entire session through global feature encoding. Specifically, the local feature sequence $\boldsymbol{Z}_\text{local}$ is input into a global feature encoder, which has the same architecture as the local encoder but with separate parameters. 
The global encoder models cross-phase dependencies to generate the global session representation:
\begin{equation}
  \boldsymbol{Z}_\text{global}=\textsf{GF-Bi-GRU}\left(\boldsymbol{Z}_\text{local}\right)\in \mathbb{R}^{2\cdot l \cdot d}
\end{equation}
In summary, this hierarchical session encoding method utilizes local feature extraction to capture short-term critical behavior patterns, and global feature extraction to model long-range semantic dependencies, achieving both fine-grained detection capability and robustness in malicious traffic analysis.

\begin{table*}[!htb]
  \centering
  \caption{Details of the new dataset.}
  \resizebox{\textwidth}{!}{
    \renewcommand\arraystretch{1.5}
    \begin{tabular}{ccccl}
\hline\hline
New Dataset              & $M$                   & Number of samples $|\mathcal{D}|$                                                                        & Data Source      & Categories of Malicious Traffic                                                                                                                                                               \\ 
\hline
\multirow{3}{*}{CIC-IDS-FS} & \multirow{3}{*}{$12$} & \multirow{3}{*}{$12\times 2000\times 2=48000$}   & CIC-IDS-2017~\cite{CIC-IDS}    & \begin{tabular}[c]{@{}l@{}}Botnet, FTP\_Patator, SSH\_Patator, Dos\_slowloris\end{tabular}                                                                                       \\
\cline{4-5}                         
                         &                     &   & CSE-CIC-IDS2018~\cite{CIC-IDS} & \begin{tabular}[c]{@{}l@{}}Botnet, Dos\_GoldenEye, Dos\_Hulk, Dos\_slowloris, DDoS-LOIC-HTTP\end{tabular}                                                                         \\
                         \cline{4-5}
                         &                     &   & CIC-DDoS-2019~\cite{43}   & \begin{tabular}[c]{@{}l@{}}MSSQL, NetBIOS, NTP, TFTP, UDP\_DDos\end{tabular}                                                                                                  \\
\hline
TON-IOT-FS                  & $6$                   & $6\times 2000\times 2=24000$      & TON-IOT~\cite{TON-IoT}         & \begin{tabular}[c]{@{}l@{}}DDoS\_HTTP, DDoS-HTTPS, injection, password, runsomware, XSS\end{tabular}                                                                        \\
\hline
\multirow{3}{*}{IDS-FS}  & \multirow{3}{*}{$23$} & \multirow{3}{*}{$23\times 2000 \times 2=92000$}  & CIC-IDS-FS         & \begin{tabular}[c]{@{}l@{}}Botnet, FTP\_Patator, SSH\_Patator, Dos\_slowloris, Dos\_GoldenEye, Dos\_Hulk,\\MSSQL, UDP\_DDos, NTP, TFTP, DDoS-LOIC-HTTP, NetBIOS\end{tabular} \\
\cline{4-5}             
                          &                     & & TON-IOT         & \begin{tabular}[c]{@{}l@{}}DDoS-HTTPS, injection, password, XSS, runsomware\end{tabular}                                                                                      \\
                          \cline{4-5}
                          &                     & & CTU-13~\cite{CTU}          & \begin{tabular}[c]{@{}l@{}}Cridex, Dridex, Geodo, Miuref, Neris, Trickbot\end{tabular}                                                                                         \\ \hline\hline
\end{tabular}}
      \label{tb: dataset}
\end{table*}

\subsection{Session Similarity Assessment}
In few-shot malicious traffic detection tasks, the limited number of samples per malicious class often prevents supervised classification methods from effectively capturing class-specific features. To address this limitation, we further design a session similarity assessment network that classifies query samples by computing their similarity to support set samples across different categories. The core idea is to enhance sample representations by fusing local similarity and global features, thereby improving classification accuracy in few-shot scenarios. Local similarity calculations capture fine-grained differences between sessions at each phase, while global feature enhancement provides comprehensive contextual information, enabling the model to leverage the full scope of query samples. This hierarchical similarity assessment method enables more accurate similarity measurements between samples and establishes robust criteria for subsequent malicious traffic detection. The architecture of the similarity assessment network is illustrated in Fig.~\ref{fig: framework}(c).

First, we compute the local similarity between sessions to enhance fine-grained matching. For the local feature sequences of a query sample (Session A) and a support sample (Session B), denoted as  $\boldsymbol{Z}^A_{\text{local}} = [\boldsymbol{z}_1^A, \boldsymbol{z}_2^A, \dots, \boldsymbol{z}_L^A]$ and $\boldsymbol{Z}^B_{\text{local}} = [\boldsymbol{z}_1^B, \boldsymbol{z}_2^B, \dots, \boldsymbol{z}_L^B]$, we compute the cosine similarity between all pairs of local features from both sessions, apply softmax normalization, and amplify differences between high and low similarity values via element-wise squaring. This yields a local similarity matrix $\boldsymbol{L} \in \mathbb{R}^{L \times L}$, formalized as follows:
\begin{equation}
  \begin{aligned}
    &\boldsymbol{L}_{ij}\leftarrow \cos\left({\boldsymbol{z}_i^A}^\top \cdot\boldsymbol{z}_j^B\right)\\
    &\boldsymbol{L}_{ij}\leftarrow\operatorname{Softmax}\left(\boldsymbol{L}_{ij}\right)=\frac{\exp \left(\boldsymbol{L}_{ij}\right)}{\sum_{a=1}^L \sum_{b=1}^L \exp \left(\boldsymbol{L}_{ab}\right)}\\
    &\boldsymbol{L}_{ij}\leftarrow \boldsymbol{L}_{ij}^2
  \end{aligned}
\end{equation}

% ...
To leverage global session information, we apply a self-attention module to enhance the global feature vectors. The self-attention mechanism captures inter-dimensional correlations within the input features, assigning importance weights to highlight critical dimensions and suppress irrelevant ones. Specifically, by employing three linear projections, the query ($\boldsymbol{q}$), key ($\boldsymbol{k}$), and value ($\boldsymbol{v}$) vectors are derived from the global feature $\boldsymbol{Z}_{\text{global}}$, followed by matrix multiplication with scaled softmax normalization to produce enhanced global features:
\begin{equation}
  \begin{aligned}
    \boldsymbol{q} = \boldsymbol{Z}_{\text{global}} \boldsymbol{W}_Q,\quad \boldsymbol{k} &= \boldsymbol{Z}_{\text{global}} \boldsymbol{W}_K,\quad \boldsymbol{v} = \boldsymbol{Z}_{\text{global}} \boldsymbol{W}_V \\
    \textsf{SAN}(\boldsymbol{Z}_{\text{global}}) &= \operatorname{Softmax}\left(\frac{\boldsymbol{q}\boldsymbol{k}^\top}{\sqrt{d_k}}\right) \boldsymbol{v}
  \end{aligned}
\end{equation}
where $\boldsymbol{W}_Q, \boldsymbol{W}_K, \boldsymbol{W}_V \in \mathbb{R}^{2 \cdot l \cdot d \times d_k}$ are projection matrices, and $d_k$ is the projection dimension.

After global feature enhancement, the network fuses the enhanced global features with the local similarity. To ensure permutation invariance, \ie the network's output remains consistent regardless of the input order of Session A and Session B, we transpose the local similarity matrix for Session B before flattening. Specifically, Session A's enhanced global features are concatenated with the flattened local similarity matrix, while Session B's enhanced global features are concatenated with the transposed and flattened local similarity matrix. These concatenated vectors are then passed through a multi-layer perceptron (MLP) to generate compact representations:
\begin{equation}
  \begin{gathered}
    \boldsymbol{Z}_A=\textsf{MLP}\left(\operatorname{Cat}\left(\operatorname{SAN}\left(\boldsymbol{Z}_\text{global}^A\right), \operatorname{ Flatten }(\boldsymbol{L})\right)\right) \\
    \boldsymbol{Z}_B=\textsf{MLP}\left(\operatorname{Cat}\left(\operatorname{SAN}\left(\boldsymbol{Z}_\text{global}^B\right), \operatorname{Flatten}\left(\boldsymbol{L}^\top\right)\right)\right)
    \end{gathered}
\end{equation}
where $\text{Cat}(\cdot)$ denotes the concatenation operation and $\text{Flatten}(\cdot)$ reshapes $\boldsymbol{L}$ from $L\times L$ to $1\times L^2$. Finally, the representations of the two sessions are concatenated and passed through another MLP, followed by a Sigmoid function to produce a similarity score in the range $[0,1]$
\begin{equation}
  \operatorname{sim}(A,B)=\operatorname{Sigmoid}\left(\operatorname{MLP}\left(\operatorname{Cat}\left(\boldsymbol{Z}_A,\boldsymbol{Z}_B\right)\right)\right)
\end{equation}

\subsection{Malicious Traffic Detection}
Finally, \model determines the traffic category of a query sample by computing its average similarity to support set samples across different classes. Specifically, for a query sample (\ie session to be detected) $q\in \mathcal{Q}$, \model first calculate its similarity to every support sample $s$ in the support set $\mathcal{S}$. It then aggregates the average similarity between the query and each class-specific support subset $\mathcal{S}_c\subset \mathcal{S}$. Finally, the query sample is classified into the category corresponding to the support subset with the highest average similarity:
\begin{equation}
  \hat{y}_q = \arg\max_{c\in\mathcal{C}}  \frac{1}{|\mathcal{S}_c|}  \sum_{s\in\mathcal{S}_c}\operatorname{sim}(q,s)
\end{equation}
where $\mathcal{S}_c$ denotes the set of support samples belonging to class $c$ within the support set $\mathcal{S}$. 
We further formulate the average similarities between the query sample and different support subsets into a logit vector $\hat{\boldsymbol{y}}_q$, where the $i$-th element represents the average similarity between the query sample and the support samples of class $i$. This logit vector is normalized via the sigmoid function and subsequently used for loss function computation during model training.

\subsection{Model Training} 
During model training, we adopt the Mean Squared Error (MSE) as the loss function, which directly measures the discrepancy between the predicted similarity scores and the ideal similarity values derived from the ground-truth labels. Specifically, for each query sample in the query set, we convert its ground-truth label into a one-hot encoded vector and compute the loss as follows:
\begin{equation}
  \mathcal{L}=\frac{1}{|\mathcal{Q}|}\sum_{i=1}^{|\mathcal{Q}|} \left\lVert \boldsymbol{y}_{q_i} -\hat{\boldsymbol{y}}_{q_i}\right\rVert^2
\end{equation}
where $|\mathcal{Q}|$ denotes the total number of query samples, and $\boldsymbol{y}_{q_i}$ represents the one-hot encoded label for query sample $q_i$, with dimensions equal to the number of classes in the support set.
Note that MSE is chosen over the commonly used cross-entropy loss for two primary reasons. First, the trainable components of \model consist only of the session feature extraction module and the similarity assessment module, whose outputs are similarity scores rather than direct class probabilities. Second, by transforming ground-truth labels into one-hot encodings, we enforce a learning objective in which the average similarity between a query sample and its true class should approach 1, while similarities with other classes should approach 0. MSE directly optimizes this objective, enabling \model to progressively increase similarity with the correct class while minimizing similarity to incorrect ones, thereby improving detection accuracy for malicious traffic.

\section{Evaluation} \label{sec:Evaluation}
\subsection{Datasets}
In the field of malicious traffic detection, several public datasets are available, such as CIC-IDS-2017~\cite{CIC-IDS}, CTU-13~\cite{CTU}, and TON-IOT~\cite{TON-IoT}. While these datasets contain a substantial volume of attack traffic, they typically offer only a limited number of attack categories, making them unsuitable for few-shot malicious traffic detection tasks. To support the training and evaluation of few-shot detection models, we reconstruct three new datasets specifically designed for few-shot malicious traffic detection by sampling and combining subsets from these existing public datasets. The details of the reconstructed datasets are reported in Table~\ref{tb: dataset}.
%
%
%
\begin{itemize}[leftmargin=10pt]
  \item \textbf{CIC-IDS-FS:} This dataset is constructed by integrating three public datasets: CIC-IDS-2017, CSE-CIC-IDS-2018~\cite{CIC-IDS}, and CIC-DDoS-2019~\cite{43}. The suffix ``-FS'' indicates its design for few-shot scenarios. CIC-IDS-2017 contains 5 days of normal traffic and 14 categories of malicious traffic, while CSE-CIC-IDS-2018 introduces 7 novel attack types. CIC-DDoS-2019 supplements diverse DDoS attack samples to address the insufficient representation of DDoS traffic in the former datasets. To resolve inconsistencies in labels (\eg redundant or conflicting attack definitions), we excludes low-quality samples through filtering, ultimately retaining 12 malicious traffic categories with clear semantics and balanced distributions.
  \item \textbf{TON-IOT-FS:} This dataset is derived from the original TON-IOT~\cite{TON-IoT} dataset, which encompasses normal and malicious traffic across a variety of IoT devices. To adapt the dataset for few-shot learning scenarios and ensure data quality, we carefully filtered the malicious traffic categories in TON-IOT. As a result, 6 representative attack types were retained to form the final version of the dataset.
  \item \textbf{IDS-FS:} To address the limited number of malicious traffic categories in the CIC-IDS-FS and TON-IOT-FS datasets, we construct a larger-scale malicious traffic dataset, IDS-FS, by merging data from multiple sources. In addition to combining CIC-IDS-FS and TON-IOT-FS, we address the scarcity of botnet traffic samples by incorporating multiple types of botnet attacks from the CTU-13 dataset (\eg IRC-based and HTTP-based botnets). These samples are integrated into a unified classification scheme through protocol alignment and label normalization. The resulting IDS-FS dataset encompasses 23 attack categories, encompassing both traditional network threats and IoT-specific attack types. This expanded class diversity significantly enhances the applicability of few-shot learning and improves the generalization capability of detection models.
\end{itemize}
For each reconstructed dataset, we randomly sample 2,000 instances for each finalized attack category from the corresponding data sources. For benign traffic, we first aggregate benign samples from all data sources, and then randomly sample a number of benign instances equal to the total number of malicious samples, ensuring a balanced distribution between benign and malicious traffic.

\subsection{Comparison Methods}

To evaluate the effectiveness and superiority of our proposed \model, we select four representative few-shot detection methods for comparison:
\begin{itemize}[leftmargin=10pt]
  \item \textbf{FCNet}~\cite{27}: It is a few-shot network traffic intrusion detection method based on meta learning. It employs a 3D convolutional network (F-Net) to extract spatio-temporal traffic features and a symmetric contrastive network (C-Net) to construct cross-sample distance metrics, enabling adaptive detection of previously unseen protocol-based attacks with only a small number of samples.
  \item \textbf{FCAD}~\cite{CIC-IDS}: It is a model-agnostic meta-learning method for few-shot class-adaptive anomaly detection. It integrates statistical feature filtering, LSTM-based autoencoders for time-series feature extraction, and a meta-learning multi-task training framework to rapidly adapt and detect unknown anomaly traffic categories with limited samples.
  \item \textbf{TF}~\cite{44}: It is a N-shot learning method for website fingerprinting attacks based on triplet network. It constructs a triplet training framework with anchor, positive, and negative samples, combined with semi-hard negative mining to optimize the feature embedding space. This enables efficient identification of target websites in Tor-encrypted traffic across varying network environments and time periods with minimal training data.
  \item \textbf{RBRN}~\cite{45}: It is an end-to-end encrypted traffic classification method based on meta learning. It incorporates a Glow-based sample generator to address data imbalance, an encoder-decoder architecture for automatic feature extraction, and a relation network to model inter-sample similarity, achieving high-efficiency unknown traffic classification and cross-dataset generalization with limited labeled samples.
\end{itemize}

\subsection{Evaluation Metrics}
To evaluate model performance across different classification tasks, we design a targeted evaluation metric system based on task characteristics and data distribution.

In binary classification scenarios (\ie detecting whether traffic is benign or malicious), we adopt Accuracy, Recall, and False Positive Rate (FPR) as evaluation metrics. Accuracy provides a global assessment of model performance by measuring overall prediction correctness. Recall focuses on the model's ability to identify positive samples (\eg malicious traffic), aiming to reduce the risk of missed detections. FPR quantifies the proportion of benign samples incorrectly classified as malicious, helping to control false alarm costs and avoid unnecessary resource consumption.

In multi-class classification scenarios (\ie detecting whether traffic is benign or belongs to a specific malicious subcategory), we adopt Precision, Recall, and F1-score as evaluation metrics. Precision measures the confidence of the model's predictions, which is especially important in scenarios requiring high-precision and fine-grained classification. F1-score, as the harmonic mean of precision and recall, balances both metrics and is particularly suitable for imbalanced class distributions. These metrics are computed per class to avoid evaluation bias caused by class dominance, thus providing a more accurate depiction of the model's performance in fine-grained malicious traffic classification.

\subsection{Experiment Settings}\label{sec: experiment_set}

\subsubsection{Common Settings}
All experiments are conducted on Ubuntu system, equipped with an AMD EPYC 9754 CPU, 60 GB of RAM, and an NVIDIA RTX 3090 GPU (24 GB video memory). \model is implemented using PyTorch version 1.9.0 and executed on CUDA version 11.1. In addition, the model's hyperparameters are set as follows: the optimizer used is Adam with a learning rate of 0.001; training is performed via 1000 episodes; each query set includes 15 samples per category; the session length $N$ is fixed at 16, and the sliding window length is set to 2. These common settings ensure reproducibility and provide a consistent basis for evaluating the performance of \model across different experiments.

\subsubsection{Binary Classification Settings}
For binary classification experiments, we evaluate \model on three constructed datasets: CIC-IDS-FS, TON-IOT-FS, and IDS-FS. In these experiments, the goal is to differentiate between benign and malicious traffic. As our few-shot learning approach is episode-based, training and test episodes are sampled separately from disjoint subsets of malicious traffic categories. In each dataset, the malicious categories are partitioned into separate training and test sets. Each few-shot episode is formulated as an ``$J$-way $K$-shot'' problem, where $J$ denotes the number of categories in the support set and $K$ is the number of samples per category. For the CIC-IDS-FS dataset, which includes 12 malicious categories, 8 malicious categories (along with an equal number of benign traffic samples) are randomly selected for the training task set, while the remaining 4 malicious categories (along with an equal number of benign traffic samples) formed the test set, from which 1,000 test episodes are generated. We conduct 5 independent experiments under both 2-way-1-shot and 2-way-5-shot settings, each with different random splits, and the final performance is obtained by averaging across these runs. Similar experimental settings are applied to the TON-IOT-FS dataset (which contains 6 malicious categories, partitioned into 4 for training and 2 for testing) and the larger IDS-FS dataset (with 23 malicious categories, where 15 are used for training and 8 for testing).

\subsubsection{Multi-class Classification Settings}
To further evaluate our model's performance in more complex scenarios, multi-class classification experiments are conducted on the IDS-FS dataset. In these experiments, we design episodes under two settings: 5-way-1-shot and 5-way-5-shot, with five independent runs for each setting. Each episode consists of samples from five categories, specifically four distinct malicious traffic categories and one benign traffic category. For each experimental run, 15 malicious categories (along with an equal number of benign traffic samples) are randomly selected to construct the training set, from which multi-class training tasks are sampled. The remaining 8 malicious categories, combined with benign traffic samples, constitute the test set, from which 1,000 test episodes are generated. This multi-class evaluation framework provides a comprehensive assessment of our model's ability to distinguish between multiple traffic classes under few-shot learning conditions.

\begin{figure*}[!htb] 
  \centering \includegraphics[width=\textwidth]{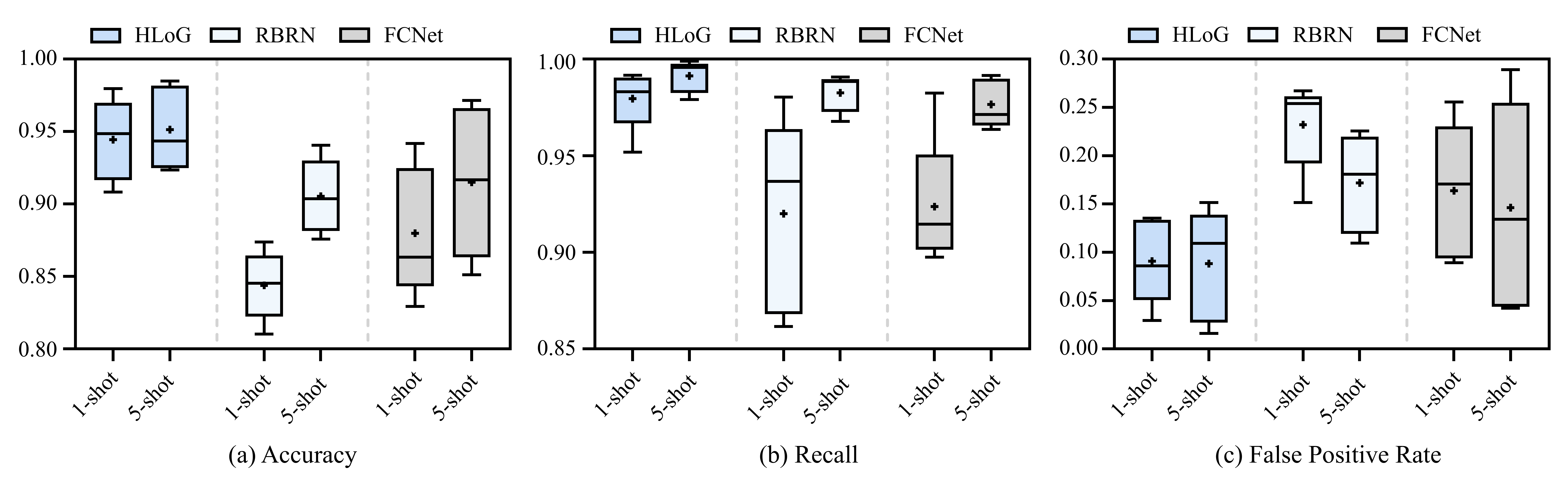} 
  \caption{Performance distribution for few-shot malicious traffic detection under binary classification settings on the IDS-FS dataset.} 
  \label{fig: box} 
\end{figure*}
\begin{figure*}[!htb] 
  \centering \includegraphics[width=0.9\textwidth]{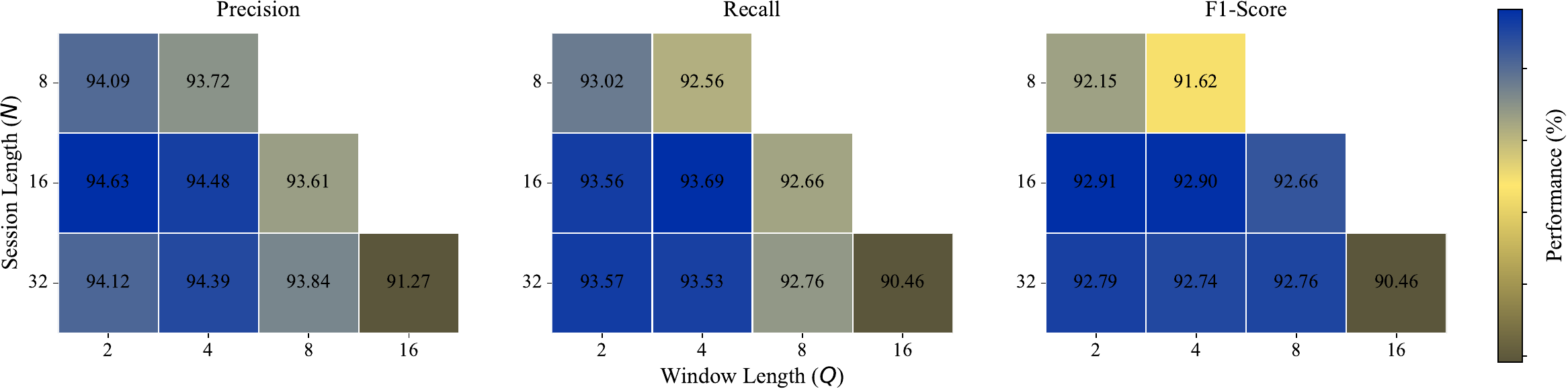} 
  \caption{Impact of session and window lengths on detection performance.} 
  \label{fig: hotmap} 
\end{figure*}

\subsection{Performance Evaluation on Malicious Traffic Detection}
Table~\ref{tab: res-ton-iot}, \ref{tab: res-cic-ids} and \ref{tab: res-ids-fs} report binary classification results for different models across three datasets. As shown, our \model demonstrates significant advantages in few-shot malicious traffic detection. Under both 2-way-1-shot and 2-way-5-shot settings, \model achieves superior average accuracy and false positive rate (FPR) compared to all baseline methods. Notably, in the 2-way-5-shot experiment on the IDS-FS dataset, \model attains a recall of 99.14\% with an FPR of 9.00\%, reflecting a balanced trade-off between detection precision and false alarms. Although FCNet achieves the highest average recall (96.78\%) in the 2-way-5-shot experiment on TON-IOT-FS, its accuracy (93.54\%) and FPR (9.70\%) lag behind \model (accuracy 96.33\%, FPR 1.76\%), highlighting our method's superior global performance. The TF method, which relies solely on packet direction features without capturing multi-dimensional session characteristics, underperforms significantly.
Moreover, we compare the performance of all methods under the multi-class classification setting. As shown in Table~\ref{tab: multi-class-ids-fs}, \model consistently outperforms all baselines in both the 5-way-1-shot and 5-way-5-shot settings.
\begin{table}[!htb]
  \caption{Few-shot malicious traffic detection results under binary classification settings on the TON-IOT-FS dataset.}
  \centering
  \resizebox{\linewidth}{!}{
      \renewcommand\arraystretch{1.2}
  \begin{tabular}{ccccccc}
  \hline\hline
  \multirow{2}{*}{Method}      & \multicolumn{3}{c}{2-way-1-shot}                & \multicolumn{3}{c}{2-way-5-shot}                \\ \cline{2-7} 
     & Accuracy            & Recall             & FPR           & Accuracy            & Recall             & FPR           \\ \hline
  TF      & 68.35          & 74.11          & 37.41         & 73.62          & 80.66          & 28.42         \\
  FCAD    & 81.67          & 86.37          & 24.03         & 87.87          & 92.17          & 17.43         \\
  FCNet   & 87.81          & 91.14          & 15.52         & 93.54          & \textbf{96.78} & 9.70          \\
  RBRN    & 78.69          & 87.13          & 29.75         & 84.51          & 93.44          & 24.41         \\
  \model & \textbf{93.96} & \textbf{92.88} & \textbf{4.95} & \textbf{96.33} & 94.43          & \textbf{1.76} \\ \hline\hline
  \end{tabular}}
  \label{tab: res-ton-iot}
\end{table}
\begin{table}[!htb]
  \caption{Few-shot malicious traffic detection results under binary classification settings on the CIC-IDS-FS dataset.}
  \centering
  \resizebox{\linewidth}{!}{
      \renewcommand\arraystretch{1.2}
  \begin{tabular}{ccccccc}
  \hline\hline
  \multirow{2}{*}{Method}  & \multicolumn{3}{c}{2-way-1-shot}                & \multicolumn{3}{c}{2-way-5-shot}                \\ \cline{2-7} 
             & Accuracy            & Recall             & FPR           & Accuracy            & Recall             & FPR           \\ \hline
  TF      & 72.45          & 78.94          & 34.04         & 76.36          & 83.17          & 31.45         \\
  FCAD    & 86.46          & 92.71          & 21.79         & 92.37          & 93.07          & 8.33          \\
  FCNet   & 92.10          & 96.98          & 12.77         & 95.08          & 96.83          & 6.68          \\
  RBRN    & 85.67          & 94.52          & 22.92         & 92.60          & 94.62          & 9.42          \\
  \model & \textbf{96.62} & \textbf{98.08} & \textbf{4.84} & \textbf{98.35} & \textbf{98.87} & \textbf{2.16} \\ \hline\hline
  \end{tabular}}
  \label{tab: res-cic-ids}
\end{table}
\begin{table}[!htb]
  \caption{Few-shot malicious traffic detection results under binary classification settings on the IDS-FS dataset.}
  \centering
  \resizebox{\linewidth}{!}{
      \renewcommand\arraystretch{1.2}
  \begin{tabular}{ccccccc}
  \hline\hline
  \multirow{2}{*}{Method}  & \multicolumn{3}{c}{2-way-1-shot}                & \multicolumn{3}{c}{2-way-5-shot}                \\ \cline{2-7} 
             & Accuracy            & Recall             & FPR           & Accuracy            & Recall             & FPR           \\ \hline
  TF      & 70.36          & 78.23          & 38.51         & 77.83          & 85.33          & 29.67         \\
  FCAD    & 83.18          & 90.03          & 23.67         & 89.91          & 95.43          & 15.61         \\
  FCNet   & 87.99          & 92.37          & 16.38         & 91.51          & 97.66          & 14.63         \\
  RBRN    & 84.39          & 92.01          & 23.22         & 90.53          & 98.26          & 17.18         \\
  \model & \textbf{94.45} & \textbf{97.95} & \textbf{9.02} & \textbf{95.06} & \textbf{99.14} & \textbf{9.00} \\ \hline\hline
  \end{tabular}}
  \label{tab: res-ids-fs}
\end{table}
\begin{table}[!htb]
  \caption{Few-shot malicious traffic detection results under multi-class classification settings on the IDS-FS dataset.}
  \centering
  \resizebox{\linewidth}{!}{
    \renewcommand\arraystretch{1.2}
  \begin{tabular}{ccccccc}
  \hline\hline
          & \multicolumn{3}{c}{5-way-1-shot}                 & \multicolumn{3}{c}{5-wat-5-shot}                             \\ \cline{2-7} 
  Method   & Precision            & Recall         & F1-score             & Precision & Recall         & F1-score             \\ \hline
  TF      & 68.36          & 67.13          & 66.87          & 72.28                      & 71.26          & 71.11          \\
  FCAD    & 85.52          & 83.00          & 79.87          & 86.81                      & 84.51          & 82.58          \\
  FCNet   & 87.82          & 86.08          & 84.36          & 90.08                      & 87.91          & 86.42          \\
  RBRN    & 86.02          & 82.85          & 80.87          & 92.32                      & 89.90          & 89.05          \\
  \model & \textbf{92.01} & \textbf{90.73} & \textbf{89.68} & \textbf{94.63}             & \textbf{93.56} & \textbf{92.91} \\ \hline\hline
  \end{tabular}}
  \label{tab: multi-class-ids-fs}
  \end{table}

Further analysis reveals that increasing the number of malicious traffic categories in the training set generally improves the average recall during testing. This trend is evident across the three datasets: TON-IOT-FS (4 classes), CIC-IDS-FS (8 classes), and IDS-FS (15 classes), suggesting that richer malicious traffic diversity enhances the model's generalization to unseen attacks. Additionally, increasing the number of support samples per class (from 1-shot to 5-shot) generally improves performance, indicating that more support examples strengthen the model's ability to capture distinguishing features of malicious traffic.
However, it is also observed that as the number of malicious classes increases, the FPR tends to rise. This is because FPR reflects the rate at which benign traffic is misclassified as malicious. Given the high diversity of benign traffic, introducing more malicious classes during training may introduce additional ambiguity in distinguishing benign samples, thereby increasing the risk of false alarms. This highlights the necessity of balancing recall and FPR in real-world applications.

To better understand performance variability, we further analyze the detection stability of \model, RBRN, and FCNet on the IDS-FS dataset using box plots, as shown in Fig.~\ref{fig: box}. The plots reveal that increasing the number of support samples not only improves the overall detection accuracy but also reduces recall variability, confirming that larger support sets contribute to both performance and robustness. Interestingly, while the average FPR tends to decrease with more support samples, its variance widens significantly. This suggests that larger support sets may amplify the uncertainty in benign traffic classification, resulting in wider FPR fluctuations.

In summary, \model demonstrates strong performance, stability, and generalization ability across multiple datasets and experimental settings, particularly excelling in achieving high recall with low FPR. The experiments also emphasize the critical role of training data diversity and support set size in enhancing few-shot malicious traffic detection.


\subsection{More analysis}
\subsubsection{Impact of Session and Window Lengths}
We further investigate the impact of different session and window lengths on model detection performance, validating our findings through 5-way-5-shot experiments on the IDS-FS dataset. As shown in Fig.~\ref{fig: hotmap}, smaller window sizes (\eg 2 or 4) yield superior detection performance. This is primarily because a smaller window allows \model to partition a session into more fine-grained phases, resulting in a local similarity matrix that provides more detailed information, which in turn enhances detection effectiveness. In contrast, larger windows (\eg 8 or 16) may over-aggregate local information, leading to a decline in performance.
Furthermore, the experimental results also indicate that longer session lengths do not necessarily yield better performance. When sessions are longer, achieving a fixed-length representation requires padding shorter sessions with additional zero entries. This redundant information can impair the model's ability to extract authentic traffic features, thereby diminishing detection accuracy. Consequently, designing session representations requires a balance between ensuring complete information and avoiding the introduction of excessive redundant or ineffective padding.

\begin{table}[!htb]
  \caption{Performance comparison of \model with different feature extraction units.}
  \centering
  \resizebox{\linewidth}{!}{
    \renewcommand\arraystretch{1.2}
  \begin{tabular}{cccc}
  \hline\hline
  \multicolumn{1}{l}{Traffic Feature Extraction Unit} & Precision            & Recall         & F1-score             \\ 
  \hline
  \model(RNN)                                          & 94.37          & 93.12          & 92.46          \\
  \model(LSTM)                                         & 94.26          & 93.18          & 92.29          \\
  \model(GRU)                                          & \textbf{94.63} & \textbf{93.56} & \textbf{92.91} \\ 
  \hline\hline
  \end{tabular}}
  \label{tab: diff-unit}
\end{table}
\begin{table}[!htb]
    \caption{Comparison of results using different methods for local similarity matrix calculation}
    \centering
    \resizebox{\linewidth}{!}{
      \renewcommand\arraystretch{1.2}
    \begin{tabular}{cccc}
    \hline\hline
    Local Similarity Measure                  & Precision            & Recall         & F1-score             \\ 
    \hline
    \model(Euclidean Distance) & 91.86          & 90.30          & 89.18          \\
    \model(Cosine Similarity)  & \textbf{94.63} & \textbf{93.56} & \textbf{92.91} \\ 
    \hline\hline
    \end{tabular}}
    \label{tab: similarity-measure}
\end{table}
\begin{table}[!htb]
    \caption{Performance comparison of similarity assessment networks under different feature combinations}
    \centering
    \resizebox{\linewidth}{!}{
      \renewcommand\arraystretch{1.2}
    \begin{tabular}{ccccc}
    \hline\hline
    Local Similarity  & Global Feature & Precision            & Recall         & F1-score             \\ \hline
    $\times$          & \checkmark          & 87.21          & 85.67          & 83.95          \\
    \checkmark       & $\times$             & 93.07          & 92.37          & 91.48          \\
    \checkmark       & \checkmark          & \textbf{94.63} & \textbf{93.56} & \textbf{92.91} \\ \hline\hline
    \end{tabular}}
    \label{fig: feature-importance}
\end{table}

\subsubsection{Impact of Feature Extraction Unit}
During session feature extraction, we employ GRU as the core sequential feature extractor. To assess its effectiveness relative to other sequence models, we further replace the GRU in \model with alternatives such as vanilla RNN and LSTM, and compare their performance under the 5-way-5-shot setting on the
IDS-FS dataset. As shown in Table~\ref{tab: diff-unit}, the performance gaps among the three models are marginal, with GRU achieving relatively superior feature extraction results.
A reasonable explanation lies in the nature of the few-shot malicious traffic detection task, where the support set contains only 1 to 5 samples. In such data-scarce settings, models with excessive parameters are prone to overfitting. GRU, with its simplified architecture, reduces model complexity and enhances generalization capability, thereby alleviating the risk of overfitting to noisy features.
Moreover, network traffic sessions typically have short sequence lengths (\eg 16 packets per session in our experimental settings), and critical discriminative patterns such as protocol handshakes and payload distributions are often concentrated in local time steps. GRU is well-suited for capturing these short-range dependencies, whereas the long-term memory capabilities of LSTM may be unnecessary or even redundant in this context.

\subsubsection{Impact of Local Similarity Measure}

During session similarity assessment, we utilize cosine similarity to measure the local similarity between session samples. Considering the availability of other similarity metrics such as Euclidean distance, we further compare their performance differences by replacing the cosine similarity component in \model. As shown in Table~\ref{tab: similarity-measure}, \model with cosine similarity significantly outperforms that with Euclidean distance in quantifying the local similarity of sessions.
A reasonable explanation is that, unlike Euclidean distance, cosine similarity focuses on the directional consistency between vectors rather than their absolute magnitudes. In malicious traffic detection, the feature vectors obtained through preprocessing are normalized, so their magnitude differences are usually small. However, the directional information among different traffic patterns can better reflect their semantic characteristics. By measuring the cosine of the angle between vectors, cosine similarity captures subtle differences in sample features while disregarding absolute scale, thereby enabling a more accurate assessment of similarity between session samples.
Moreover, Euclidean distance is sensitive to scale, so when feature vectors contain noise or have inconsistent magnitudes, its performance may be significantly disrupted. In contrast, the scale invariance of cosine similarity renders it more robust under such conditions, ultimately enhancing detection performance.

\subsubsection{Feature Importance on Session Similarity Assessment}
Finally, we further analyze the importance of different features employed in the session similarity assessment process, as shown in Table~\ref{fig: feature-importance}. It can be observed that when global features are removed (\ie only local features are used for similarity evaluation), the performance of \model slightly declines; however, when local similarity is removed (i.e., only global features are employed), the performance drops dramatically. This demonstrates that local features play a crucial role in measuring session similarity.
A reasonable explanation is that local features are adept at capturing fine-grained, key behavioral patterns within a session. These patterns may manifest as anomalies only within specific phases and serve as direct signals of malicious activities. Although global features can encapsulate the overall trend of an entire session, the aggregation process may smooth out or dilute these vital local anomalies. Consequently, relying exclusively on global features for similarity evaluation might cause the model to overlook subtle yet discriminative local information, leading to a substantial decline in performance. In contrast, preserving local features enables the model to more sensitively detect nuanced differences between sessions, thereby significantly enhancing detection accuracy.

\section{Conclusion} \label{sec:Conclusion}
Experimental results across multiple reconstructed benchmark datasets demonstrate that the proposed \model framework achieves state-of-the-art performance in few-shot malicious traffic detection. Compared with existing methods, \model consistently achieves higher accuracy and recall, while significantly reducing false positive rates in both binary and multi-class classification settings. The model shows strong generalization capability to unseen attack types and exhibits stable performance across varying few-shot configurations, validating its effectiveness in realistic low-data scenarios.

Despite these promising results, several directions remain for future improvement. First, adaptive mechanisms for handling imbalanced or noisy traffic data could further enhance robustness. Second, integrating online or continual learning strategies may improve the model's responsiveness to newly emerging threats in evolving network environments. Lastly, applying the hierarchical local-global feature learning framework to broader cybersecurity tasks such as anomaly detection or encrypted traffic classification offers exciting opportunities for future research.

\bibliographystyle{IEEEtran} 
\bibliography{mybib,IEEEabrv}

\begin{IEEEbiography}[{\includegraphics[width=1in,height=1.25in,clip,keepaspectratio]{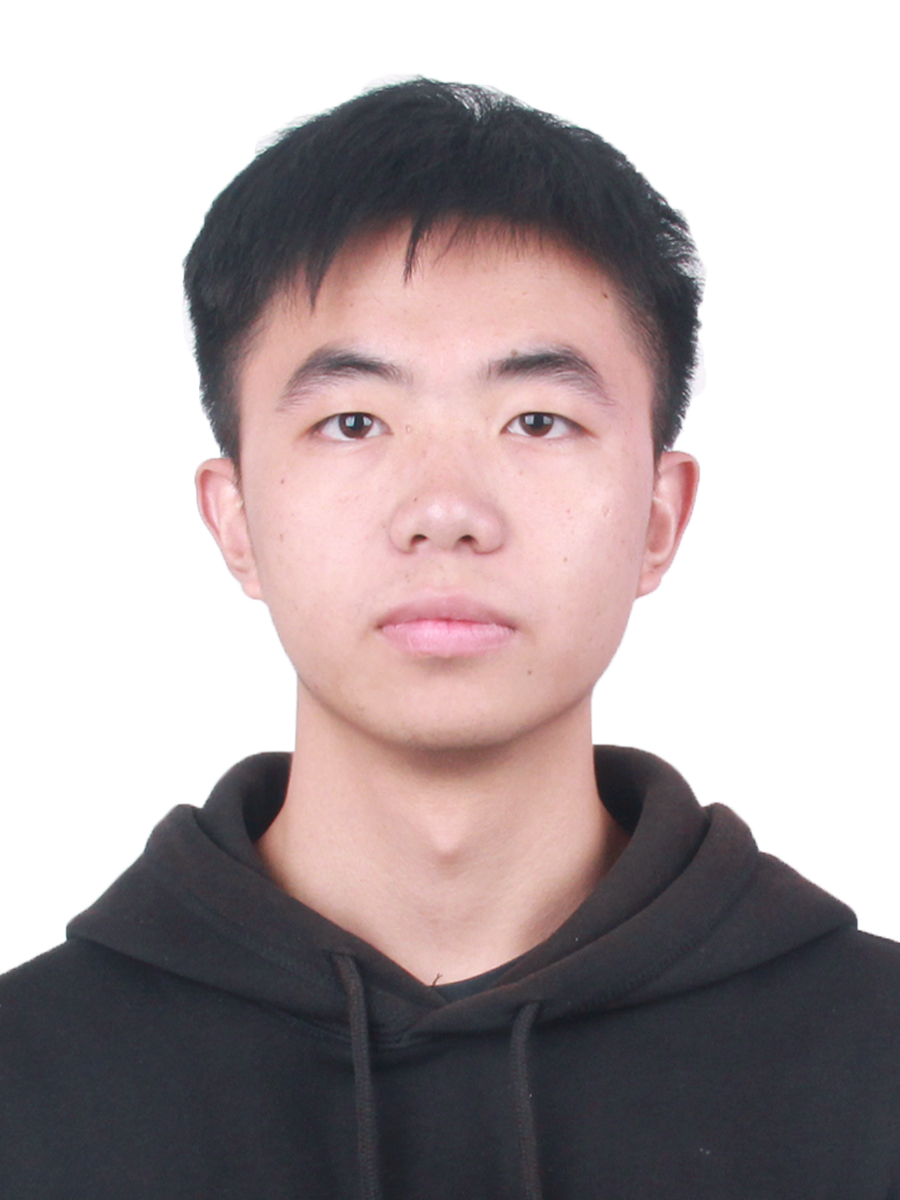}}]{Songtao Peng}
 received the B.S. degree in Huzhou College, Huzhou, China, in 2020. He is currently pursuing the Ph.D degree in control science and engineering with the Institute of Cyberspace Security, Zhejiang University of Technology, Hangzhou, China. His research interests include complex network, social network analysis, anomaly detection.
\end{IEEEbiography}

\begin{IEEEbiography}[{\includegraphics[width=1in,height=1.25in,clip,keepaspectratio]{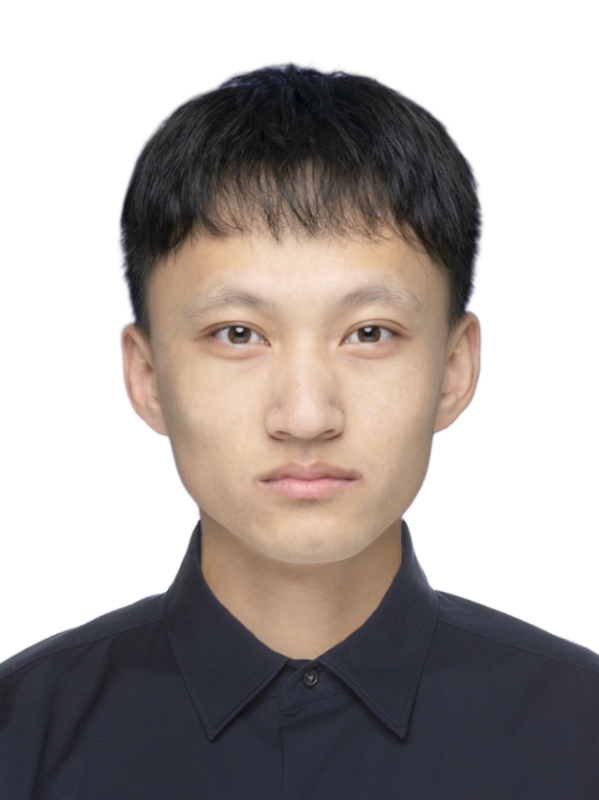}}]{Lei Wang}
   received the M.S. degree in control engineering with the Institute of Cyberspace Security, Zhejiang University of Technology, Hangzhou, China, in 2024. His research interests include deep learning, few-shot learning, and malicious traffic detection.
 \end{IEEEbiography}

 \begin{IEEEbiography}[{\includegraphics[width=1in,height=1.25in,clip,keepaspectratio]{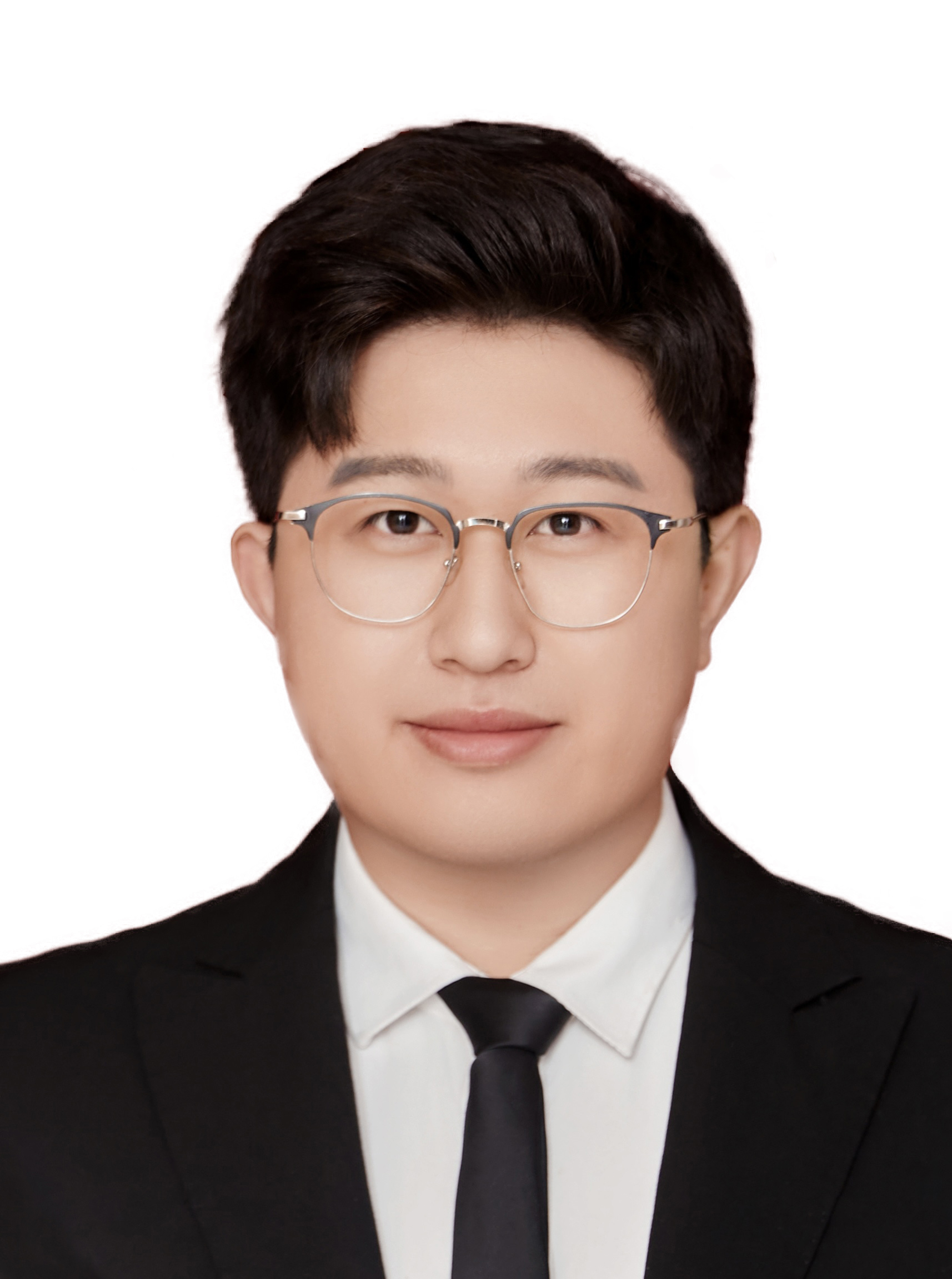}}]{Wu Shuai}
   received the B.S. degree in Zhejiang University of Science and Technology , Hangzhou, China, in 2022. He is currently pursuing the MS degree in control engineering at Zhejiang University of Technology, Hangzhou, China. His current research interests include deep learning incremental learning and malicious traffic detection.
\end{IEEEbiography}

\begin{IEEEbiography}[{\includegraphics[width=1in,height=1.25in,clip,keepaspectratio]{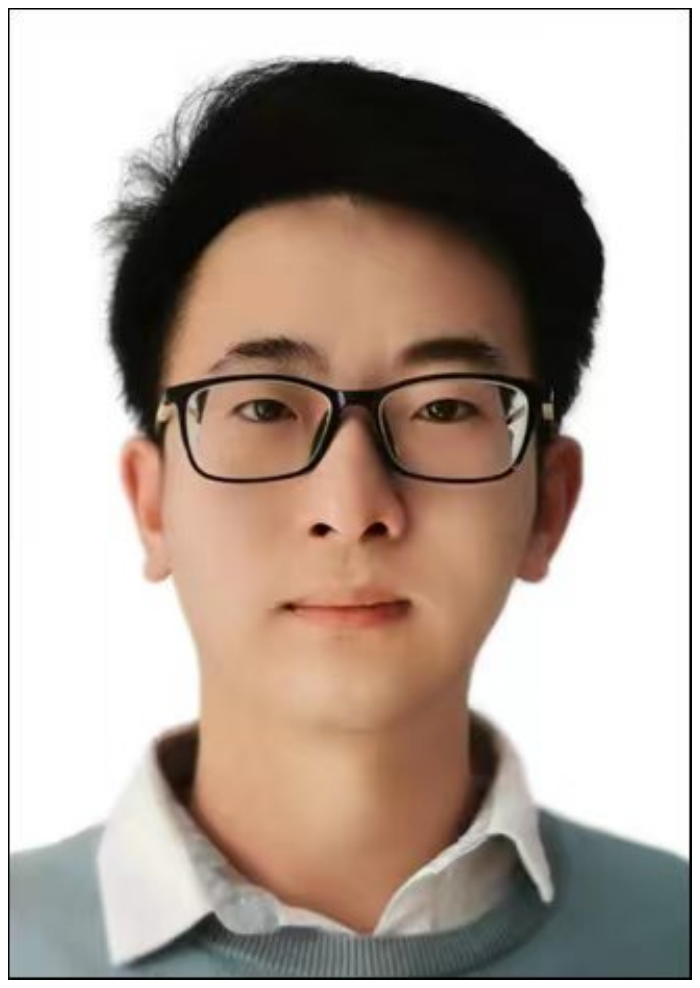}}]{Hao Song}
 received the B.S. degree in Communication Engineering from Nanjing Institute of Technology, Nanjing, China, in 2021. He is currently pursuing his master's degree at the Institute of Cyberspace Security, Zhejiang University of Technology, China. His current research interests include cyberspace security and network intrusion detection.
\end{IEEEbiography}

\begin{IEEEbiography}[{\includegraphics[width=1in,height=1.25in,clip,keepaspectratio]{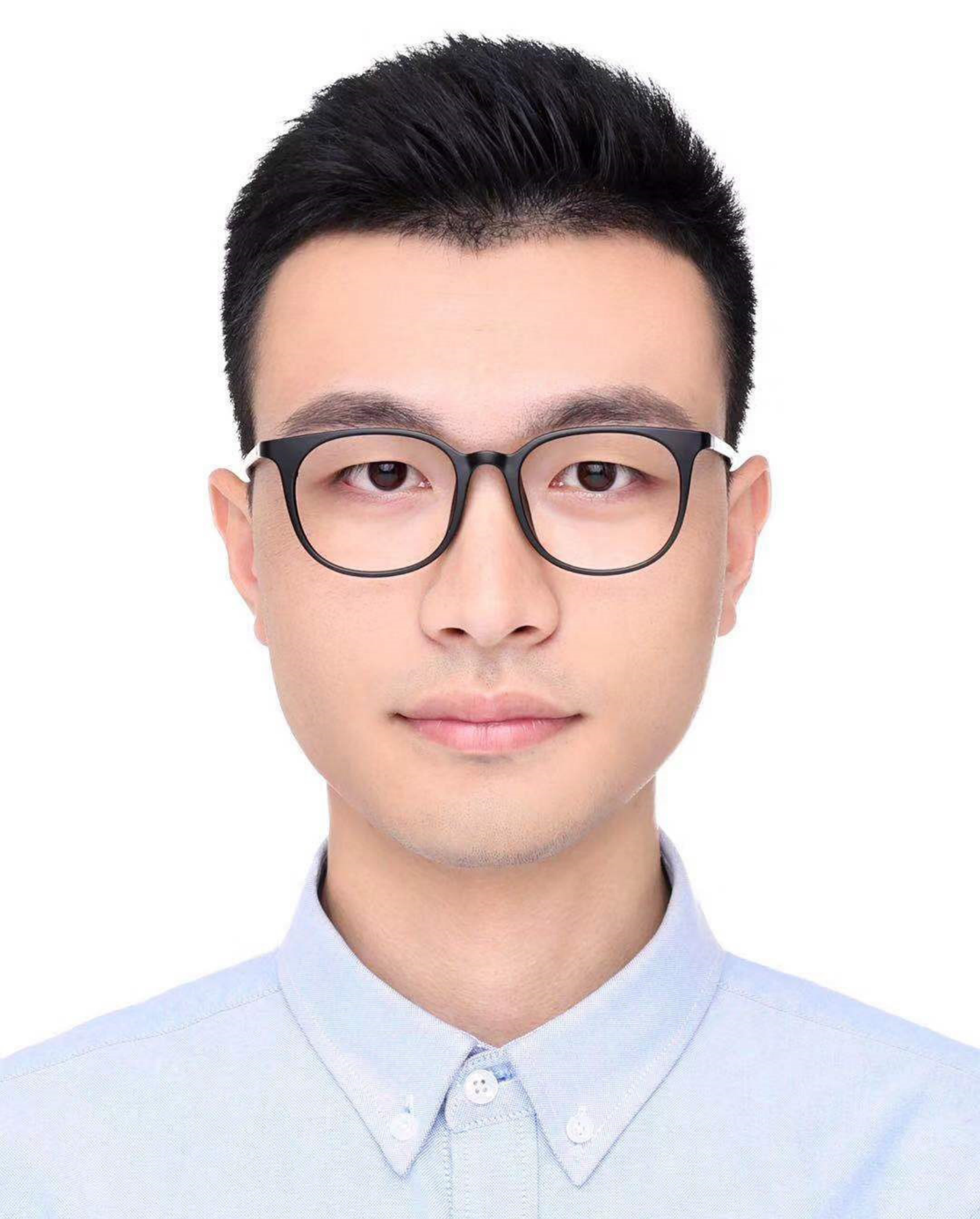}}]{Jiajun Zhou}
	received the Ph.D degree in control theory and engineering from Zhejiang University of Technology, Hangzhou, China, in 2023. He is currently a Research Assistant Professor and Postdoctoral Fellow with the Institute of Cyberspace Security, Zhejiang University of Technology. His current research interests include graph data mining, cyberspace security and data management.
\end{IEEEbiography}

\begin{IEEEbiography}[{\includegraphics[width=1in,height=1.25in,clip,keepaspectratio]{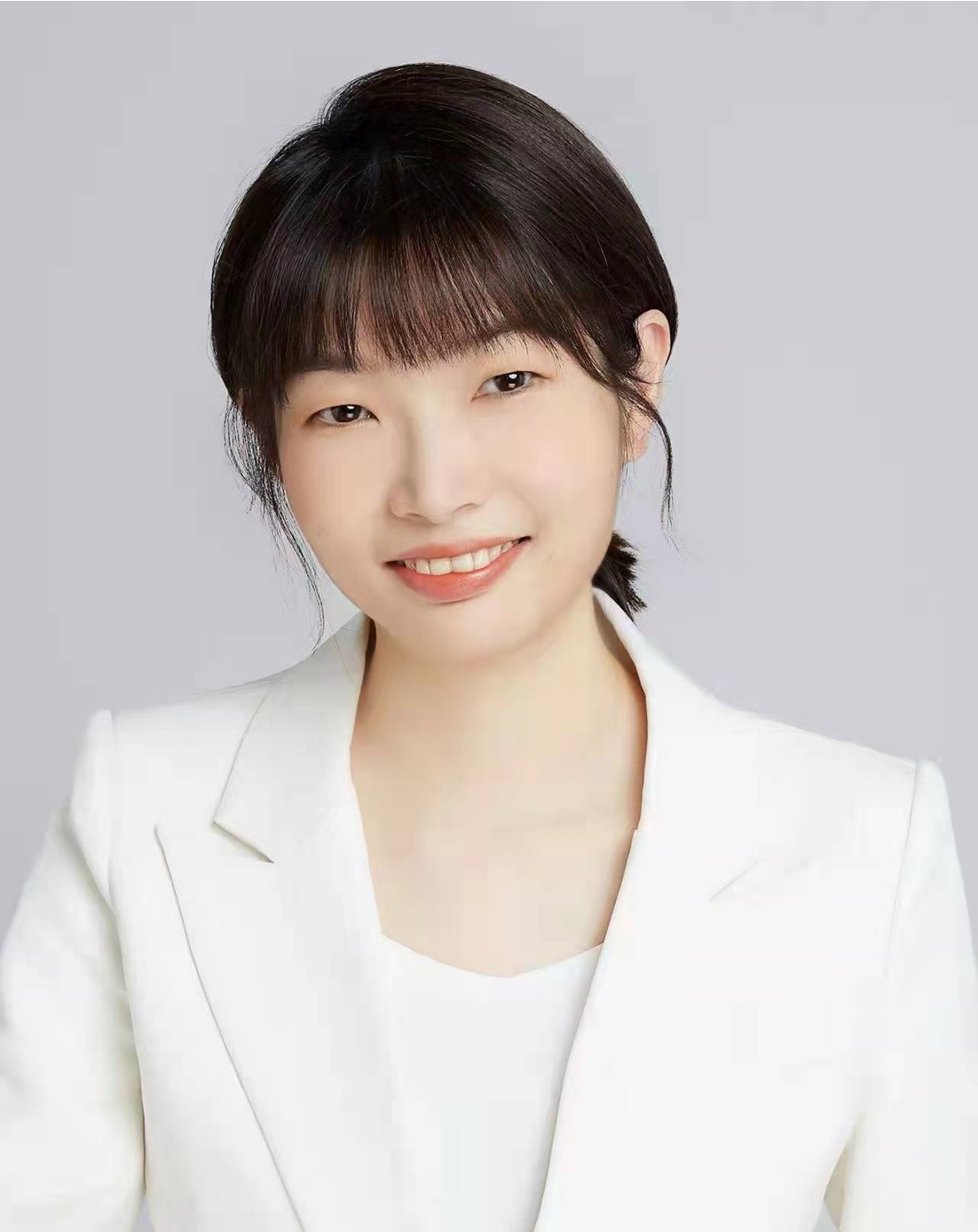}}]{Shanqing Yu}
	received the M.S. degree from the School of Computer Engineering and Science, Shanghai University, China, in 2008 and received the M.S. degree from the Graduate School of Information, Production and Systems, Waseda University, Japan, in 2008, and the Ph.D. degree, in 2011, respectively. She is currently a Lecturer at the Institute of Cyberspace Security and the College of Information Engineering, Zhejiang University of Technology, Hangzhou, China. Her research interests cover intelligent computation and data mining.
\end{IEEEbiography}

\begin{IEEEbiography}[{\includegraphics[width=1in,height=1.25in,clip,keepaspectratio]{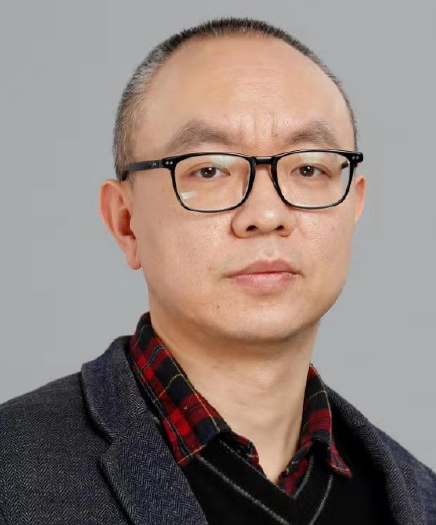}}]{Qi Xuan}(M'18) received the BS and PhD degrees in control theory and engineering from Zhejiang University, Hangzhou, China, in 2003 and 2008, respectively. He was a Post-Doctoral Researcher with the Department of Information Science and Electronic Engineering, Zhejiang University, from 2008 to 2010, respectively, and a Research Assistant with the Department of Electronic Engineering, City University of Hong Kong, Hong Kong, in 2010 and 2017. From 2012 to 2014, he was a Post-Doctoral Fellow with the Department of Computer Science, University of California at Davis, CA, USA. He is a senior member of the IEEE and is currently a Professor with the Institute of Cyberspace Security, College of Information Engineering, Zhejiang University of Technology, Hangzhou, China. His current research interests include network science, graph data mining, cyberspace security, machine learning, and computer vision.
\end{IEEEbiography}


\end{document}